\documentclass[useAMS,usenatbib]{mnras_tex_edited} 

\usepackage{amsmath,array,amsfonts}
\usepackage{amssymb}
\usepackage{graphicx}
\usepackage{tabularx}
\usepackage{placeins}
\usepackage{float}
\usepackage{algorithm}
\usepackage[noend]{algpseudocode}
\usepackage{natbib}
\usepackage{pstricks}
\usepackage{verbatim}
\usepackage{subfigure}
\usepackage{booktabs}
\usepackage{siunitx,booktabs,multirow, dcolumn}
\usepackage{hyperref}

\newcommand{\ltsima} {$\; \buildrel < \over \sim \;$}
\newcommand{\gtsima} {$\; \buildrel > \over \sim \;$}
\newcommand{\lta} {\lower.5ex\hbox{\ltsima}}
\newcommand{\gta} {\lower.5ex\hbox{\gtsima}}

\newcolumntype{d}[1]{D{.}{.}{#1} }

\def\f{\frac}

\def\ln{\mathrm{ln}}
\def\exp{\mathrm{exp}}

\def\data{\mathbf{d}}

\def\mmax{m_\mathrm{max}}
\def\mp{m_\mathrm{p}}
\def\mc{m_\mathrm{c}}
\def\mt{m_\mathrm{T}}
\def\dee{\mathrm{d}}
\def\msol{M_\odot}
\def\btheta{\boldsymbol{\theta}}
\def\rhoo{\rho_\mathrm{sat}}

\begin{document}

\title[Evidence for a maximum neutron star mass cut-off and constraints on the EoS]{Evidence for a maximum mass cut-off in the neutron star mass distribution and constraints on the equation of state}

\author[J. Alsing, H. O. Silva, E. Berti]
{\parbox{\textwidth}{Justin Alsing$^1$\thanks{e-mail:  jalsing@flatironinstitute.org},
Hector O. Silva$^{2,3}$, 
Emanuele Berti$^{3,4}$
}\vspace{0.4cm}\\
\parbox{\textwidth}{$^1$ Center for Computational Astrophysics, Flatiron Institute, 162 5th Avenue, New York, NY 10010, USA\\
$^2$ eXtreme Gravity Institute, Department of Physics, Montana State University, Bozeman, MT 59717 USA \\
$^3$ Department of Physics and Astronomy, The University of Mississippi, University, MS 38677-1848, USA \\
$^4$ CENTRA, Departamento de F\'{i}sica, Instituto Superior T\'{e}cnico, Universidade de Lisboa, Avenida Rovisco Pais 1, 1049 Lisboa, Portugal}}
\date{Accepted ;  Received ; in original form }

\maketitle

\begin{abstract}
We infer the mass distribution of neutron stars in binary systems using a flexible Gaussian mixture model and use Bayesian model selection to explore evidence for multi-modality and a sharp cut-off in the mass distribution. We find overwhelming evidence for a bimodal distribution, in agreement with previous literature, and report for the first time positive evidence for a sharp cut-off at a maximum neutron star mass. We measure the maximum mass to be $2.0\msol < \mmax < 2.2\msol$ (68\%), $2.0\msol < \mmax < 2.6\msol$ (90\%), and evidence for a cut-off is robust against the choice of model for the mass distribution and to removing the most extreme (highest mass) neutron stars from the dataset. If this sharp cut-off is interpreted as the maximum stable neutron star mass allowed by the equation of state of dense matter, our measurement puts constraints on the equation of state. For a set of realistic equations of state that support $>2\msol$ neutron stars, our inference of $\mmax$ is able to distinguish between models at odds ratios of up to $12:1$, whilst under a flexible piecewise polytropic equation of state model our maximum mass measurement improves constraints on the pressure at $3-7\times$ the nuclear saturation density by $\sim 30-50\%$ compared to simply requiring $\mmax > 2\msol$. We obtain a lower bound on the maximum sound speed attained inside the neutron star of $c_s^\mathrm{max} > 0.63c$ (99.8\%), ruling out $c_s^\mathrm{max} < c/\sqrt{3}$ at high significance. Our constraints on the maximum neutron star mass strengthen the case for neutron star-neutron star mergers as the primary source of short gamma-ray bursts.
\end{abstract}
\section{Introduction}
\label{sec:intro}
The distribution of neutron star (NS) masses encodes a wealth of information about NS physics: NS formation channels, compact binary evolution via mass accretion, and the equation of state (EoS) of matter at ultra-high densities in the NS interior all leave distinct observable signatures on the NS mass distribution. A large enough set of NS mass measurements required for detailed study of the mass distribution has only become available relatively recently \citep{Valentim2011, Ozel2012, Kiziltan2013, Antoniadis2016} thanks mainly to a sustained and ongoing effort in radio timing of pulsars in binary systems, with earlier studies being limited to challengingly small sample sizes \citep{Joss1976, Finn1994, Thorsett1999, Schwab2010}.

Previous studies have reported strong evidence for a bimodal NS mass distribution \citep{Valentim2011, Ozel2012, Kiziltan2013, Antoniadis2016}, with one peak at $\sim 1.3\msol$ and a second peak around $\sim1.5-1.7\msol$. This is expected physically, since different formation and evolution channels result in NS masses clustered around different values: see \citet{Horvath2016} and references therein for a recent review.

In addition to being multimodal, it is possible that the mass distribution has a sharp cut-off at the highest stable mass supported by the EoS of NS matter, $\mmax$. The observation of a sharp cut-off and hence the determination of $\mmax$ would put important constraints on the EoS, and it is also interesting as it delineates the low-mass limit of stellar mass black holes \citep{Fryer2001}. Recent measurement of NSs with masses close to $2\msol$ \citep{Antoniadis2013, Demorest2010, Fonseca2016} has already put significant constraints on the EoS, and the requirement that $>2\msol$ NSs are supported has become one of the cornerstones of observational constraints on the nuclear EoS at ultra-high densities.

Whilst observations of $2\msol$ NS puts a robust lower bound on $\mmax$, obtaining a strongly constraining upper limit has proven more elusive. From a theoretical perspective, requiring that the EoS satisfies causality and our knowledge of nuclear matter at low densities provides a loose upper limit of $\mmax < 2.9\msol$, as shown by \citet{Kalogera1996}, following an earlier calculation by \citet{Rhoades1974}. More recently, \citet{Lawrence2015} and \citet{Fryer2015} obtained an upper bound on $\mmax$ from analysis of short gamma-ray bursts (GRBs). They argued that NS mergers only produce short GRBs if the core of the remnant collapses quickly to a black hole, and this is only possible for EoSs with relatively low $\mmax$. For EoSs with too high a maximum mass, only a tiny fraction of NS-NS mergers are able to produce GRBs; therefore if GRBs are primarily produced in NS-NS mergers then this would require a merger rate much higher than canonical values, severely stretching our understanding of binary evolution. Based on this argument \citet{Fryer2015} find $\mmax < 2.2-2.3\msol$. \citet{Lawrence2015} similarly find $\mmax < 2.2-2.5\msol$ ($\mmax < 2.2\msol$ assuming the rotation of the remnant is limited by mass shedding, and $\mmax < 2.5\msol$ in the limiting case where the remnant has no angular momentum).

In a recent study of the mass distribution of millisecond pulsars, \citet{Antoniadis2016} considered the possibility of a sharp truncation in the mass distribution and obtained a posterior distribution for $\mmax$ peaked at $\simeq 2.1\msol$ but with reasonably large uncertainties, and from their small sample size (32 millisecond pulsars) significant evidence for or against a sharp cut-off could not be firmly established.

In this work we study the distribution of NS masses using all available NS mass measurements and explore the case for a sharp cut-off in the mass distribution due to the EoS. While previous studies of the mass distribution have split the NS sample by rotation period or other characteristics that are intended to separate out different accretion histories \citep{Valentim2011, Ozel2012, Kiziltan2013, Antoniadis2016}, here we take a different approach and model the mass distribution for the whole population together using a flexible Gaussian mixture model. The mixture model has the advantage that it can naturally elicit subpopulations populating distinct modes of the distribution, and is flexible enough to capture highly non-Gaussian distributions; this allows us to analyze all the NS mass data together to obtain the strongest possible constraints on $\mmax$.

The structure of this paper is as follows. In \S \ref{sec:mass_likelihoods} we describe the NS mass measurements. In \S \ref{sec:model} we describe the truncated Gaussian mixture models used to model the mass distribution, and the Bayesian parameter inference and model selection approaches are covered in \S \ref{sec:param_inference}. The results for the inferred mass distribution are discussed in \S \ref{sec:inferred_mass_dist}, and the evidence for and constraints on a cut-off in the NS mass distribution are presented in \S \ref{sec:mmax_constraints}. In \S \ref{sec:eos_constraints} we explore the implications of the inferred $\mmax$ on the NS EoS, computing constraints on a set of realistic EoSs in \S \ref{sec:numerical_eos} and on a parameterized piecewise-polytropic EoS in \S \ref{sec:polytropic_eos}, including constraints on the maximum sound speed in NS matter. Discussion and conclusions are in \S \ref{sec:conclusions}.
\section{Neutron star mass measurements}
\label{sec:mass_likelihoods}
In this section we describe the NS mass measurements that constitute the dataset used in this work, summarized in Table \ref{tab:mass_data}\footnote{A database of mass measurements is maintained at {\sc{https://stellarcollapse.org}}, which was useful in compiling the data in Table \ref{tab:mass_data}.}. For systems with measurements of two or more post-Keplerian parameters (or alternatively mass ratio or companion mass) we assume Gaussian mass likelihoods taken from the references in Table \ref{tab:mass_data}, and similarly for x-ray/optical observations. For systems where only the total mass or mass ratio are measured, we combine these measurements with the observed mass function to form the likelihoods given in Eqs.  \eqref{total_mass_likelihood}--\eqref{q_likelihood} below. For a recent review of NS mass measurements see \citet{Ozel2016}.

\subsection{Pulsars with radio timing}
Radio timing of pulsars in binary systems yields precise measurements of the Keplerian orbital parameters, in particular, the orbital period $P_b$, eccentricity $e$ and projected semi-major axis $x_p$ of the pulsar's orbit. The orbital period and projected semi-major axis together determine the mass function $f$,
\begin{align}
f = \left(\f{2\pi}{P_b}\right)^2\f{x_p^3}{G} = \f{\mc^3\sin^3i}{(\mp+\mc)^2},
\end{align}
where $\mp$ and $\mc$ are the pulsar and companion mass, and $i$ is the orbital inclination. Whilst the mass function is sensitive to the component masses, with three degenerate unknowns additional constraints are required to determine the component masses of the system.

For sufficiently compact binaries, it's also possible to measure post-Keplerian parameters describing relativistic effects on the orbital motion, that are also functions of the component masses and orbital inclination. In particular, the periastron precession $\dot{\omega}$, Einstein delay $\gamma$, shape and range of the Shapiro delay $s$ and $r$, and the period derivative $\dot{P}_b$ due to gravitational wave damping are all observable and sensitive to the component masses: see e.g. \citet{Stairs2003} and references therein. Measurement of two or more post-Keplerian parameters along with the mass function breaks the degeneracies and leads to the highest precision measurements of the component masses $\mp$ and $\mc$.

If the projected semi-major axis of the companion's orbit can also be determined, either by radio timing if it is also a pulsar, or by phase-resolved optical spectroscopy if it is optically bright, one can also determine the mass ratio of the binary $q$,
\begin{align}
q = \f{\mc}{\mp}=\f{x_p}{x_c}.
\end{align}
In cases where the companion is a main sequence star or white dwarf, the spectrum of the companion also contains information about its composition, that in turn provides an independent constraint on the companion mass $\mc$. Although these companion mass measurements are stellar-model dependent, mappings between spectral properties and mass have reached a satisfactory level of sophistication for accurate (and robust) mass determination \citep{Tremblay2013,Althaus2013,Istrate2014,Tremblay2015}.

Systems where the mass function and two or more additional constraints have been measured (either post-Keplerian parameters, $q$ or $\mc$) typically yield precise pulsar mass measurements with (close to) Gaussian uncertainties. For these systems we will assume Gaussian mass likelihoods $P(\data |\mp)$, taking the mean and variance reported in the relevant radio timing analysis paper as given in Table \ref{tab:mass_data}.

In some cases, the only additional constraint available is a measurement of the periastron precession $\dot{\omega}$, that effectively provides a measurement of the total binary mass $\mt = \mp + \mc$. In these cases the pulsar mass can still be constrained, but the likelihood for $\mp$ resulting from the under-constrained measurements is highly non-Gaussian, and this must be properly accounted for. We construct the pulsar-mass likelihood from the measured total mass $\hat{m}_T$ and mass function $\hat{f}$ as follows:
\begin{align}
\label{total_mass_likelihood}
P(\data | \mp) &\propto \iint P(\hat{m}_\mathrm{T}, \hat{f} | \mp, \mt, i)P(\mt)P(i)\dee i\,\dee \mt\nonumber \\
&\propto \iint P(\hat{m}_\mathrm{T} | \mt)P(\hat{f} | f(\mp, \mt, i)) \nonumber \\
&\quad\times P(\mt)P(i)\dee i\,\dee \mt \nonumber \\
&\propto \iint \exp\left[-\f{(\mt - \hat{m}_\mathrm{T})^2}{2\sigma_{\mt}^2}\right] \nonumber \\
&\quad\times\delta\left(f(\mp, \mt, i) - \hat{f}\right)\sin i\dee i\, \dee \mt\nonumber \\
&=\int \f{\exp\left[-\f{(\mt - \hat{m}_\mathrm{T})^2}{2\sigma_{\mt}^2}\right]\dee \mt}{\f{3(\mt-\mp)^3}{\mt^2}\left[\f{\hat{f}\mt^2}{(\mt-\mp)^3}\right]^\f{1}{3}\left[1-\f{\hat{f}^{2/3}\mt^{4/3}}{(\mt-\mp)^2}\right]^\f{1}{2}},
\end{align}
where in the second line we assume the mass function and total mass measurements are independent, in the third line we assume Gaussian uncertainties on $\mt$, negligible uncertainties on $\hat{f}$\footnote{Uncertainties on the mass function from radio timing are typically very small, $\lesssim 0.1\%$ or less, so the delta-function approximation is well justified.} and uniform priors over $\cos i$ and $\mt$, and in the final line we analytically integrate out the delta function. The resulting pulsar mass likelihood function is given by the final line above.

Similarly, there are some cases where the only measurements available are the mass function $\hat{f}$ and the mass ratio $\hat{q}$ from phase-resolved optical spectroscopy. These under-constrained systems also result in highly non-Gaussian pulsar mass likelihoods. In a similar vein to Eq. \eqref{total_mass_likelihood}, for these systems we can derive a likelihood:
\begin{align}
\label{q_likelihood}
P(\data | \mp) &\propto \iint P(\hat{q}, \hat{f} | \mp, q, i)\dee i\,\dee q \nonumber \\
&\propto \iint P(\hat{q} | q)P(\hat{f} | f(\mp, q, i))P(q)P(i)\dee i\,\dee q \nonumber \\
&\propto \iint \exp\left[-\f{(q - \hat{q})^2}{2\sigma_q^2}\right]\delta\left(f(\mp, q, i) - \hat{f}\right) \sin i\dee i\,\dee q \nonumber \\
&=\int \f{\exp\left[-\f{(q - \hat{q})^2}{2\sigma_q^2}\right]\dee q}{\f{3\mp}{q(q+1)^2}\left[\f{\hat{f}q(q+1)^2}{\mp}\right]^\f{1}{3}\left[1-\left(\f{\hat{f}q(q+1)^2}{\mp}\right)^\f{2}{3}\right]^\f{1}{2}},
\end{align}
where similarly in the second line we assume the mass function and mass ratio measurements are independent, in the third line we assume Gaussian uncertainties on $q$, negligible uncertainties on $\hat{f}$ and uniform priors over $\cos i$ and $q$\footnote{A uniform prior on $q$ might not be the most appropriate choice given knowledge of the nature of the two objects, but since the measurement uncertainties on $q$ are typically $\lesssim 1\%$, the Gaussian likelihood for $q$ should be sufficiently sharply peaked to make this highly insensitive to alternative prior choices.}, and in the final line we analytically integrate out the delta function.

For systems where only $\mt$ or $q$ were measured in addition to the mass function we assume pulsar mass likelihoods given by Eqs. \eqref{total_mass_likelihood} and \eqref{q_likelihood}, respectively.

\subsection{X-ray/optical mass measurements}
It is also possible to measure masses of NSs with high and low stellar mass companions using x-ray and optical observations.

For NSs with high-mass companions, eclipsing x-ray binaries where the companion blocks the x-rays from the pulsar during part of the orbit can yield mass measurements. X-ray observations of the pulsar give the Keplerian parameters $P_b$, $e$, $x_p$ and the time and longitude of periastron, as well as the duration of the eclipse. Optical observations of the companion allow determination of its velocity amplitude, projected rotational velocity and amplitude of ellipsoidal variation. With these measurements in hand, it is possible to solve for the parameters of the binary system, including the NS mass. For NSs with low-mass companions, observations of thermonuclear x-ray bursts can provide measurements of their masses and radii. Typically, mass constraints from x-ray and optical observations are less precise than radio timing constraints and may be subject to systematic biases: see \citet{Ozel2012, Falanga2015, Ozel2016} and references therein. Nonetheless, since all of these measurements are unlikely to be systematically biased in the same direction and make up a minority fraction of the dataset, we include them in our analysis (performing a sensitivity test to removing these data in \S \ref{sec:mmax_constraints}). For x-ray/optically determined masses, we assume Gaussian mass likelihoods taken from the relevant literature (see Table \ref{tab:mass_data}).

The full NS mass dataset and associated literature is summarized in Table \ref{tab:mass_data}.
\begin{table*}
\centering
\scalebox{0.95}{
\begin{tabularx}{\textwidth}{ccccccc}
\toprule
name & type & f [$\msol$] & $\mt$ [$\msol$] & q & $\mp$ [$\msol$] & reference\tabularnewline
\midrule
4U1700-377 & x-ray/optical &  &  &  & 1.96$\pm$0.19 & \citet{Falanga2015}\tabularnewline
Cyg X-2 & x-ray/optical &  &  &  & 1.71$\pm$0.21 & \citet{Casares2009}\tabularnewline
SMC X-1 & x-ray/optical &  &  &  & 1.21$\pm$0.12 & \citet{Falanga2015}\tabularnewline
Cen X-3 & x-ray/optical &  &  &  & 1.57$\pm$0.16 & \citet{Falanga2015}\tabularnewline
XTE J2123-058 & x-ray/optical &  &  &  & 1.53$\pm$0.42 & \citet{Gelino2002}\tabularnewline
4U 1822-371 & x-ray/optical &  &  &  & 1.96$\pm$0.36 & \citet{MunozDarias2005}\tabularnewline
OAO 1657-415 & x-ray/optical &  &  &  & 1.74$\pm$0.3 & \citet{Falanga2015}\tabularnewline
J013236.7+303228 & x-ray/optical &  &  &  & 2.0$\pm$0.4 & \citet{Bhalerao2012}\tabularnewline
Vela X-1 & x-ray/optical &  &  &  & 2.12$\pm$0.16 & \citet{Falanga2015}\tabularnewline
4U1538-522 & x-ray/optical &  &  &  & 1.02$\pm$0.17 & \citet{Falanga2015}\tabularnewline
LMC X-4 & x-ray/optical &  &  &  & 1.57$\pm$0.11 & \citet{Falanga2015}\tabularnewline
Her X-1 & x-ray/optical &  &  &  & 1.073$\pm$0.36 & \citet{Rawls2011}\tabularnewline
2S 0921-630 & x-ray/optical &  &  &  & 1.44$\pm$0.1 & \citet{Steeghs2007}\tabularnewline
EXO 1722-363 & x-ray/optical &  &  &  & 1.91$\pm$0.45 & \citet{Falanga2015}\tabularnewline
SAX J1802.7-2017 & x-ray/optical &  &  &  & 1.57$\pm$0.25 & \citet{Falanga2015}\tabularnewline
XTE J1855-026 & x-ray/optical &  &  &  & 1.41$\pm$0.24 & \citet{Falanga2015}\tabularnewline
B1957+20 & x-ray/optical &  $5\times10^{-6}$ &   & 69.2$\pm$0.8 & & \citet{Kerkwijk2011}\tabularnewline
J1311-3430 & x-ray/optical & $3\times10^{-7}$ & & 175$\pm$3 & & \citet{Romani2012}\tabularnewline
J1740-5350 & x-ray/optical &  0.002644  &   & 5.85$\pm$0.13 & & \citet{Ferraro2003}\tabularnewline
J1816+4510 & x-ray/optical &  0.0017607 &   &  9.54$\pm$0.21 & & \citet{Kaplan2013}\tabularnewline
J1723-2837 & x-ray/optical &  0.005221 &   & 3.45$\pm$0.02 & & \citet{vanStaden2016}\tabularnewline
J0453+1559 & NS-NS &  &  &  & 1.559$\pm$0.004 & \citet{Martinez2015}\tabularnewline
J0453+1559 comp. & NS-NS &  &  &  & 1.174$\pm$0.004 & \citet{Martinez2015}\tabularnewline
J1906+0746 & NS-NS &  &  &  & 1.291 $\pm$0.011 & \citet{vanLeeuwen2014}\tabularnewline
J1906+0746 comp. & NS-NS &  &  &  & 1.322$\pm$0.011 & \citet{vanLeeuwen2014}\tabularnewline
B1534+12 & NS-NS &  &  &  & 1.3332$\pm$0.0010 & \citet{Fonseca2014}\tabularnewline
B1534+12 comp. & NS-NS &  &  &  & 1.3452$\pm$0.0010 & \citet{Fonseca2014}\tabularnewline
B1913+16 & NS-NS &  &  &  & 1.4398$\pm$0.0002 & \citet{Weisberg2010}\tabularnewline
B1913+16 comp. & NS-NS &  &  &  & 1.3886$\pm$0.0002 & \citet{Weisberg2010}\tabularnewline
B2127+11C & NS-NS &  &  &  & 1.358$\pm$0.010 & \citet{Jacoby2006}\tabularnewline
B2127+11C comp. & NS-NS &  &  &  & 1.354$\pm$0.010 & \citet{Jacoby2006}\tabularnewline
J0737-3039A & NS-NS &  &  &  & 1.3381$\pm$0.0007 & \citet{Kramer2006}\tabularnewline
J0737-3039B & NS-NS &  &  &  & 1.2489$\pm$0.0007 & \citet{Kramer2006}\tabularnewline
J1756-2251 & NS-NS &  &  &  & 1.312$\pm$0.017 & \citet{Ferdman2014}\tabularnewline
J1756-2251 comp. & NS-NS &  &  &  & 1.258$\pm$0.017 & \citet{Ferdman2014}\tabularnewline
J1807-2500B & NS-NS &  &  &  & 1.3655$\pm$0.0021 & \citet{Lynch2011}\tabularnewline
J1807-2500B comp. & NS-NS &  &  &  & 1.2064$\pm$0.0020 & \citet{Lynch2011}\tabularnewline
J1913+1102 & NS-NS & 0.136344 & 2.875$\pm$0.014 &  &  & \citet{Lazarus2016}\tabularnewline
J2045+3633 & NS-WD & & & & 1.33$\pm$0.3 & \citet{Berezina2017}\tabularnewline
J2053+4650 & NS-WD & & & & 1.40$\pm$0.21 & \citet{Berezina2017}\tabularnewline
J1713+0747 & NS-WD &  &  &  & 1.35$\pm$0.07 & \citet{Arzoumanian2017}\tabularnewline
B1855+09 & NS-WD &  &  &  & 1.37$\pm$0.13 & \citet{Arzoumanian2017}\tabularnewline
J0751+1807 & NS-WD &  &  &  & 1.72$\pm$0.07 & \citet{Desvignes2016}\tabularnewline
J1141-6545 & NS-WD &  &  &  & 1.27$\pm$0.01 & \citet{Bhat2008}\tabularnewline
J1738+0333 & NS-WD &  &  &  & 1.47$\pm$0.07 & \citet{Antoniadis2012}\tabularnewline
J1614-2230 & NS-WD &  &  &  & 1.908$\pm$0.016 & \citet{Arzoumanian2017}\tabularnewline
J0348+0432 & NS-WD &  &  &  & 2.01$\pm$0.04 & \citet{Antoniadis2013}\tabularnewline
J2222-0137 & NS-WD &  &  &  & 1.76$\pm$0.06 & \citet{Cognard2017}\tabularnewline
J2234+0611 & NS-WD &  &  &  & 1.393$\pm$0.013 & \citet{Stovall2016}\tabularnewline
J1949+3106 & NS-WD &  &  &  & 1.47$\pm$0.43 & \citet{Deneva2012}\tabularnewline
J1012+5307 & NS-WD &  &  &  & 1.83$\pm$0.11 & \citet{Antoniadis2016}\tabularnewline
J0437-4715 & NS-WD &  &  &  & 1.44$\pm$0.07 & \citet{Reardon2016}\tabularnewline
J1909-3744 & NS-WD &  &  &  & 1.48$\pm$0.03 & \citet{Arzoumanian2017}\tabularnewline
J1802-2124 & NS-WD &  &  &  & 1.24$\pm$0.11 & \citet{Ferdman2010}\tabularnewline
J1911-5958A & NS-WD &  &  &  & 1.34$\pm$0.08 & \citet{Bassa2006}\tabularnewline
J2043+1711 & NS-WD &  &  &  & 1.38$\pm$0.13 & \citet{Arzoumanian2017}\tabularnewline
J0337+1715 & NS-WD &  &  &  & 1.4378$\pm$0.0013 & \citet{Ransom2014}\tabularnewline
J1946+3417 & NS-WD &  &  &  & 1.828$\pm$0.022 & \citet{Barr2016}\tabularnewline
J1918-0642 & NS-WD &  &  &  & 1.29$\pm$0.1 & \citet{Arzoumanian2017}\tabularnewline
J1600-3053 & NS-WD &  &  &  & 2.3$\pm$0.7 & \citet{Arzoumanian2017}\tabularnewline
J0024-7204H & NS-WD & 0.001927 & 1.665$\pm$0.007 &  &  & \citet{Freire2017}\tabularnewline
J0514-4002A & NS-WD & 0.14549547 & 2.453$\pm$0.014 &  &  & \citet{Freire2007a}\tabularnewline
J0621+1002 & NS-WD & 0.027026849 & 2.32$\pm$0.08 &  &  & \citet{Splaver2002}\tabularnewline
B1516+02B & NS-WD & 0.000646723 & 2.29$\pm$0.17 &  &  & \citet{Freire2008a}\tabularnewline
J1748-2021B & NS-WD & 0.0002266235 & 2.92$\pm$0.20 &  &  & \citet{Freire2008}\tabularnewline
J1748-2446I & NS-WD & 0.003658 & 2.17$\pm$0.02 &  &  & \citet{Ransom2005}\tabularnewline
J1748-2446J & NS-WD & 0.013066 & 2.20$\pm$0.04 &  &  & \citet{Ransom2005}\tabularnewline
B1802-07 & NS-WD & 0.00945034 & 1.62$\pm$0.07 &  &  & \citet{Thorsett1999}\tabularnewline
J1824-2452C & NS-WD & 0.006553 & 1.616$\pm$0.007 &  &  & \citet{Freire2008}\tabularnewline
B2303+46 & NS-WD & 0.246261924525 & 2.64$\pm$0.05 &  &  & \citet{Thorsett1999}\tabularnewline
J1750-37A & NS-WD & 0.0518649 & 1.97$\pm$0.15 &  &  & \citet{Freire2008}\tabularnewline
J0045-7319 & NS-MS &  &  &  & 1.58$\pm$0.34 & \citet{Nice2003}\tabularnewline
J1023+0038 & NS-MS &  &  &  & 1.71$\pm$0.16 & \citet{Deller2012}\tabularnewline
J1903+0327 & NS-MS &  &  &  & 1.666$\pm$0.01 & \citet{Arzoumanian2017}\tabularnewline
\bottomrule
\end{tabularx}}
\caption{NS mass measurements ($74$ in total). For systems with two or more post-Keplerian parameters measured, and x-ray/optical mass measurements, we take measured pulsar masses at face value with Gaussian uncertainties. For systems where only the mass function and total mass or mass ratio are measured, we use these data to construct the mass likelihoods given in Eqs. \eqref{total_mass_likelihood} and \eqref{q_likelihood}.}
\label{tab:mass_data}
\end{table*}

\section{Model for the neutron star mass distribution}
\label{sec:model}
We want a flexible parameterized model for the mass distribution that can allow for a number of key features: multiple modes coming from subpopulations with distinct formation channels and accretion histories, non-Gaussian (e.g. skewed) modes, and the possibility of a sharp truncation at some maximum mass. To this end, we model the NS mass distribution as an $n$-component Gaussian mixture model with a sharp cut-off at some maximum mass $\mmax$. The truncated Gaussian mixture model has all of the required features; multiple Gaussian components can capture both distinct peaks and skewed individual modes, and the possibility of a sharp cut-off can be included as a free parameter. 

We consider the model space covered by varying the number of Gaussian components $n$, with two scenarios regarding the cut-off: either a sharp cut-off fixed at $\mmax = 2.9\msol$, or keeping $\mmax < 2.9\msol$ as an additional free parameter. Exploring this model space allows us to compare models of varying complexity using Bayesian model comparison and explicitly assess the evidence for multiple modes and skewness, and to quantify evidence for a sharp cut-off at $\mmax < 2.9\msol$.

The $n$-component mixture model for the mass distribution is given by
\begin{align}
\label{mixture}
P(\mp | \btheta) = \sum_{i=1}^n r_i \mathcal{N}(\mp | \mu_i, \sigma_i)\Theta(\mp - \mmax)/\Phi_i
\end{align}
for the pulsar mass $\mp$, where $\mu_i$, $\sigma_i$ and $r_i$ denote the mean, standard deviation and relative weight of the $i$-th Gaussian component, $\mathcal{N}$ denotes the Gaussian density and $\Theta$ denotes the Heaviside function. The normalization constants $\Phi_i \equiv \Phi(\mu_i, \sigma_i; m_\mathrm{min}, \mmax)$ are integrals over the Gaussian components (over the allowed NS mass range):
\begin{align}
\Phi(\mu, \sigma; m_\mathrm{min}, \mmax) = \int_{m_\mathrm{min}}^{\mmax} \mathcal{N}(x | \mu, \sigma)\dee x.
\end{align}
To keep the distribution normalized to unity, the weights are constrained to sum to one, $\sum_{i=1}^n r_i = 1$. The full set of model parameters is hence:
\begin{align}
\btheta = \{\mu_1, \mu_2, \dots, \mu_n, \sigma_1, \sigma_2, \dots, \sigma_n, r_1, r_2, \dots, r_{n}, \mmax\}
\end{align} 
We consider two cases: one with $\mmax$ fixed at $\mmax=2.9\msol$, and one with $\mmax < 2.9\msol$ kept as a free model parameter. The following uniform priors are assumed for the model parameters: $\mu_i \in [0.9, 2.9]$, $\sigma_i \in [0.01, 2]$, $\mmax \in [1.9, 2.9]$, and a flat Dirichlet prior over the weights $\{r_i\}$\footnote{The flat Dirichlet prior $\mathbf{r}\sim\mathrm{Dir}(n; 1)$ is a uniform prior over the $(n-1)$-simplex defined by $0< r_i < 1$ $\forall i$ and $\sum_{i=1}^n r_i = 1$, ensuring that the mixture distribution Eq. \eqref{mixture} is proper and normalized to unity.}. We impose the additional constraint that the component means are ordered $(\mu_1 < \mu_2 < \dots < \mu_n)$ so that the Gaussian components are distinguishable. We assume a minimum NS mass of $m_\mathrm{min}=0.9\msol$ throughout. The limits of the uniform priors over $\{\mu_i,\,\sigma_i,\,\mmax\}$ are carefully chosen to be broad enough to not truncate the resulting posteriors (over the range of allowed parameter values). However, choice of uniform priors is still somewhat subjective.
\section{Bayesian inference of the neutron star mass distribution}
\label{sec:param_inference}
\subsection{Parameter inference}
The goal is to infer the model parameters of the NS mass distribution from the measured masses in a Bayesian framework. The joint posterior for the masses $\mp$ and mass-distribution model parameters $\btheta$ given the data (for all NSs $\data = \{\data^i\}$) is given by Bayes' theorem:
\begin{align}
P(\btheta, \{\mp^i\} | \data = \{\data^i\}) &\propto P(\btheta)P(\data | \btheta, \{\mp^i\}) \nonumber \\
&\propto P(\btheta)\prod_{i=1}^N P(\data^i | \mp^i)P(\mp^i | \btheta),
\end{align}
where in the second line we have exploited the conditional independence of the data on the parameters $\btheta$ once the masses $\{\mp^i\}$ are specified, and the assumed independence of the mass data for each NS. The posterior for $\btheta$ marginalized over the individual masses is hence given by
\begin{align}
\label{theta_posterior}
P(\btheta | \data) \propto P(\btheta)\prod_{i=1}^N \int P(\data^i | \mp^i)P(\mp^i | \btheta) d\mp^i.
\end{align}
Since the mass-distribution model considered here is a truncated Gaussian mixture, the integrals on the right hand side of Eq. \eqref{theta_posterior} have simple closed-form solutions for Gaussian mass-likelihoods, and are otherwise cheap to compute numerically for the remaining non-Gaussian likelihoods, i.e. Eqs. \eqref{total_mass_likelihood} and \eqref{q_likelihood}. We explicitly marginalize over all of the individual masses either analytically or numerically, and reconstruct the posterior $P(\btheta | \data)$ by sampling from Eq. \eqref{theta_posterior} using nested sampling with \textsc{multinest} \citep{Feroz2008, Feroz2009, Buchner2014}. Nested sampling has the advantage that it generates samples from the posterior distribution and simultaneously computes the Bayesian evidence integrals needed for performing Bayesian model selection (as required in \S \ref{sec:model_selection}, see below).
\subsection{Bayesian model selection}
\label{sec:model_selection}
In addition to inferring the model parameters of the mass-distribution models considered, we also want to be able to compare their relative merit and determine which model is preferred by the data. Bayesian model selection -- computing the odds ratio between two models -- provides a principled framework for model comparison that naturally takes into account Occam's razor, i.e., penalizing more complicated models where both models fit the data. 

The odds ratio between two models $\mathcal{M}_A$ and $\mathcal{M}_B$ is given by 
\begin{align}
\mathcal{O}_{AB} = \f{P(\mathcal{M}_A | \data)}{P(\mathcal{M}_B | \data)} = \f{P(\data | \mathcal{M}_A)}{P(\data | \mathcal{M}_B)}\f{P(\mathcal{M}_A)}{P(\mathcal{M}_B)},
\end{align}
where $\mathcal{Z} = P(\data | \mathcal{M})$ is the Bayesian evidence (or marginal likelihood) for a given model $\mathcal{M}$ given data $\data$, and the prior odds ratio $P(\mathcal{M}_A)/P(\mathcal{M}_B)$ defines our prior relative belief in model $A$ over model $B$. If we are \emph{a priori} agnostic about the two models, the prior odds is unity and the odds ratio reduces to the Bayes factor:
\begin{align}
\mathcal{K}_{AB} = \f{P(\data| \mathcal{M}_A)}{P(\data | \mathcal{M}_B)},
\end{align} 
where $\mathcal{K}_{AB} > 1$ indicates that model $A$ is preferred by the data over model $B$, and vice versa for $\mathcal{K}_{AB} < 1$. In this study we follow the scheme of \citet{Kass1995} for interpreting the Bayes factor: $2\ln\mathcal{K}_{AB} < 0$ implies no support for $A$ over $B$, $0 < 2\ln\mathcal{K}_{AB} < 2$ support for $A$ ``worth not more than a bare mention'', $2 < 2\ln\mathcal{K}_{AB} < 6$ positive support for $A$, $6< 2\ln\mathcal{K}_{AB} < 10$ strong support, and $2\ln\mathcal{K}_{AB} > 10$ very strong support.

We compute the Bayesian evidences $\mathcal{Z}$ for the models considered using \textsc{multinest} nested sampling \citep{Feroz2008, Feroz2009, Buchner2014}.
\section{Results: the inferred neutron star mass distribution}
\label{sec:inferred_mass_dist}
The Bayesian evidences for the space of (truncated) Gaussian mixture models are given in Table \ref{tab:evidences}. The models with two and three components are preferred by the data compared to the models with one and four Gaussian components. The single-component Gaussian is strongly disfavored (with Bayes factors of $2\ln\mathcal{K} \gtrsim 10$) when compared against the two and three component models, indicating very strong evidence against $n=1$. The four-component model is also disfavored although less strongly, with Bayes factors of $2\ln\mathcal{K} \gtrsim 2$ compared to the two and three component models. Comparing the two and three component models, the two component model is modestly preferred in all cases, but the difference in their Bayesian evidences is not large enough to make a strong preference for either model.

In all cases $n=1,\dots,4$, the model with the maximum NS mass as an additional free parameter is preferred. In all cases $n\geq2$, the models with free $\mmax$ are preferred over those with fixed $\mmax$ with Bayes factors $2\ln\mathcal{K}>3$: there is positive support for a sharp cut-off in the NS mass distribution at $\mmax < 2.9\msol$. This is the first major result of this paper.

\begin{figure}
\centering
\includegraphics[width = 9.5cm]{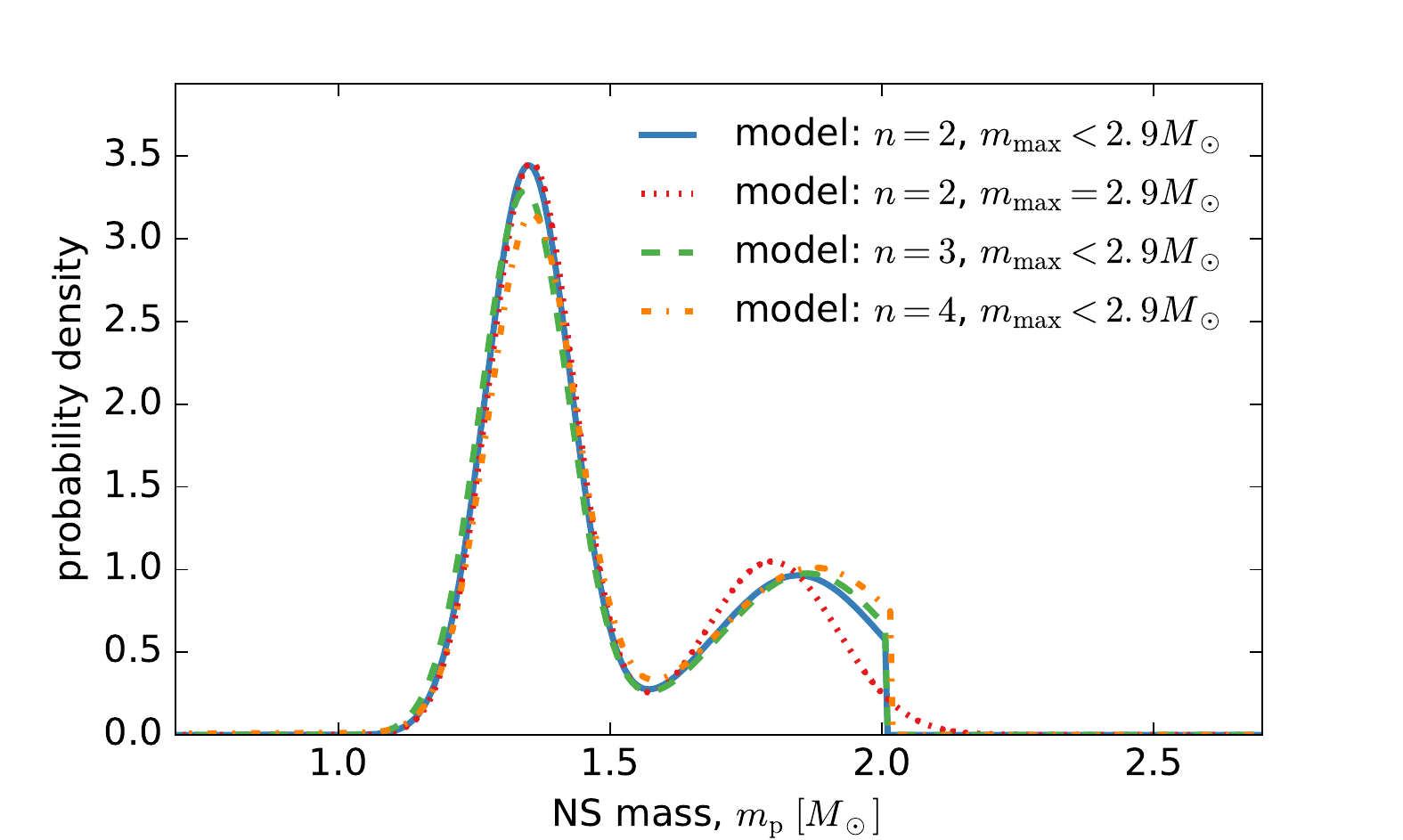}
\caption{Comparison of maximum a posteriori (MAP) NS mass distributions under different model assumptions: $n = 2$ Gaussian components with $\mmax < 2.9\msol$ (preferred model; blue-solid), $n = 2$ Gaussian components with $\mmax = 2.9\msol$ (red-dashed), $n = 3$ Gaussian components with $\mmax < 2.9\msol$ (green-dashed), $n = 4$ Gaussian components with $\mmax < 2.9\msol$ (orange-dashed).}
\label{fig:mass_dist_fid}
\end{figure}

Fig. \ref{fig:mass_dist_fid} shows the maximum \emph{a posteriori} (MAP) mass distributions for the four models that are most preferred by the data; the $n=2,3,4$ models with a sharp cut-off at $\mmax < 2.9\msol$, and the $n=2$ model with the cut-off fixed out at $\mmax=2.9\msol$ for comparison. The introduction of additional Gaussian components above $n=2$ modifies the shape of the second peak, but does not introduce an additional independent mode, even though this would be perfectly allowed under the model. The convergence of all models with $n\geq 2$ to a bimodal distribution provides overwhelming support for a bimodal NS mass distribution, with no evidence for an additional distinct peak in the distribution or separation of the lower mass peak into two narrow components, as suggested in \citet{Schwab2010}. This is consistent with recent literature \citep{Valentim2011, Ozel2012, Kiziltan2013, Antoniadis2016}.

\begin{figure}
\centering
\includegraphics[width = 9.5cm]{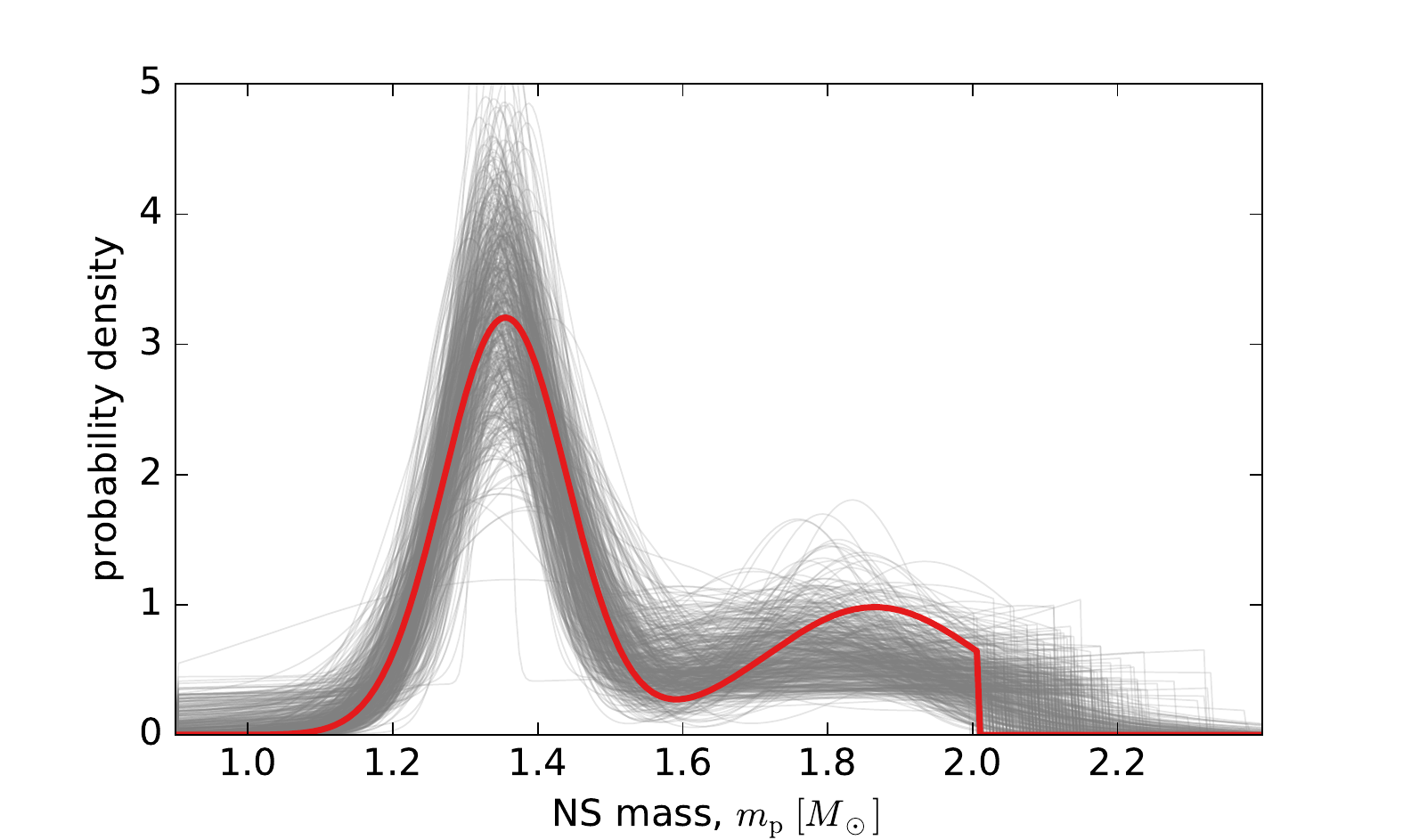}
\caption{Maximum a posteriori (MAP) NS mass distribution (red) with $1000$ independent posterior samples to give a visual guide for the uncertainties, under the considered model that is most preferred by the data; the $n=2$ component Gaussian mixture with a sharp cut-off $\mmax < 2.9\msol$.}
\label{fig:mass_dist_fid2}
\end{figure}

Fig. \ref{fig:mass_dist_fid2} shows the MAP mass distribution for the $n=2$ model with free $\mmax$ with $1000$ independent posterior samples plotted over the top to give a visual impression of the uncertainties on the shape of the distribution. As the most preferred model, we take this as our fiducial model moving forward.

The MAP values and 68\% credible regions for the preferred $n=2$ component models are given in Table \ref{tab:map_parameters}.
\begin{table}
\centering
\begin{tabularx}{0.45\textwidth}{ccc}
\toprule
 model: & $\mmax = 2.9\msol$ & $\mmax < 2.9\msol$ \tabularnewline
\midrule
$n = 1$ components & -35.0 & -34.8 \tabularnewline
$n = 2$ components & {\bf -25.8} & {\bf -22.7} \tabularnewline
$n = 3$ components & -27.3 & {\bf -23.9} \tabularnewline
$n = 4$ components & -30.4 & {\bf -25.9} \tabularnewline
\bottomrule
\end{tabularx}
\caption{Log Bayesian evidences $2\ln(\mathcal{Z})$ for the set truncated Gaussian mixture models considered for the NS mass distribution, covering $n=1\dots4$ Gaussian components, each with either fixed $\mmax = 2.9\msol$ or keeping $\mmax < 2.9\msol$ as an additional free parameter. The four preferred models are highlighted in boldface. We perform model selection by comparing $2\ln\mathcal{K}_{AB}=2\ln\mathcal{Z}_A-2\ln\mathcal{Z}_B$ to the scale of \citet{Kass1995} (see \S \ref{sec:model_selection}).}
\label{tab:evidences}
\end{table}
\begin{table*}
\begin{center}
\begin{tabularx}{\textwidth}{ccccccc}
\toprule
model &  $\mu_1$ & $\mu_2$ & $\sigma_1$ & $\sigma_2$ & $r_1$ & $\mmax$ \tabularnewline
\midrule
$n=2$ components, $\mmax < 2.9\msol$ & $1.34 ^{+ 0.03 }_{- 0.02 }$ & $1.80 ^{+ 0.15 }_{- 0.18 }$ & $0.07 ^{+ 0.02 }_{- 0.02 }$ & $0.21 ^{+ 0.18 }_{- 0.14 }$ & $0.65 ^{+ 0.08 }_{- 0.15 }$ & $2.12 ^{+ 0.09 }_{- 0.12 }$\tabularnewline

$n=2$ components, $\mmax = 2.9\msol$ & $1.34 ^{+ 0.02 }_{- 0.02 }$ & $1.78 ^{+ 0.07 }_{- 0.09 }$ & $0.07 ^{+ 0.03 }_{- 0.03 }$ & $0.12 ^{+ 0.09 }_{- 0.03 }$ & $0.66 ^{+ 0.09 }_{- 0.09 }$ & - \tabularnewline
\bottomrule
\end{tabularx}
\caption{Maximum a posteriori (MAP) values and 68\% credible intervals of the 1-d marginal posteriors for the preferred $n=2$ component Gaussian mixture model with free $\mmax < 2.9\msol$ and with $\mmax$ fixed at $2.9\msol$. All dimensional quantities are in units of solar masses.}
\label{tab:map_parameters}
\end{center}
\end{table*}
\subsection{Constraints on $\mmax$ from the neutron star mass distribution}
\label{sec:mmax_constraints}
\begin{figure}
\centering
\includegraphics[width = 9.5cm]{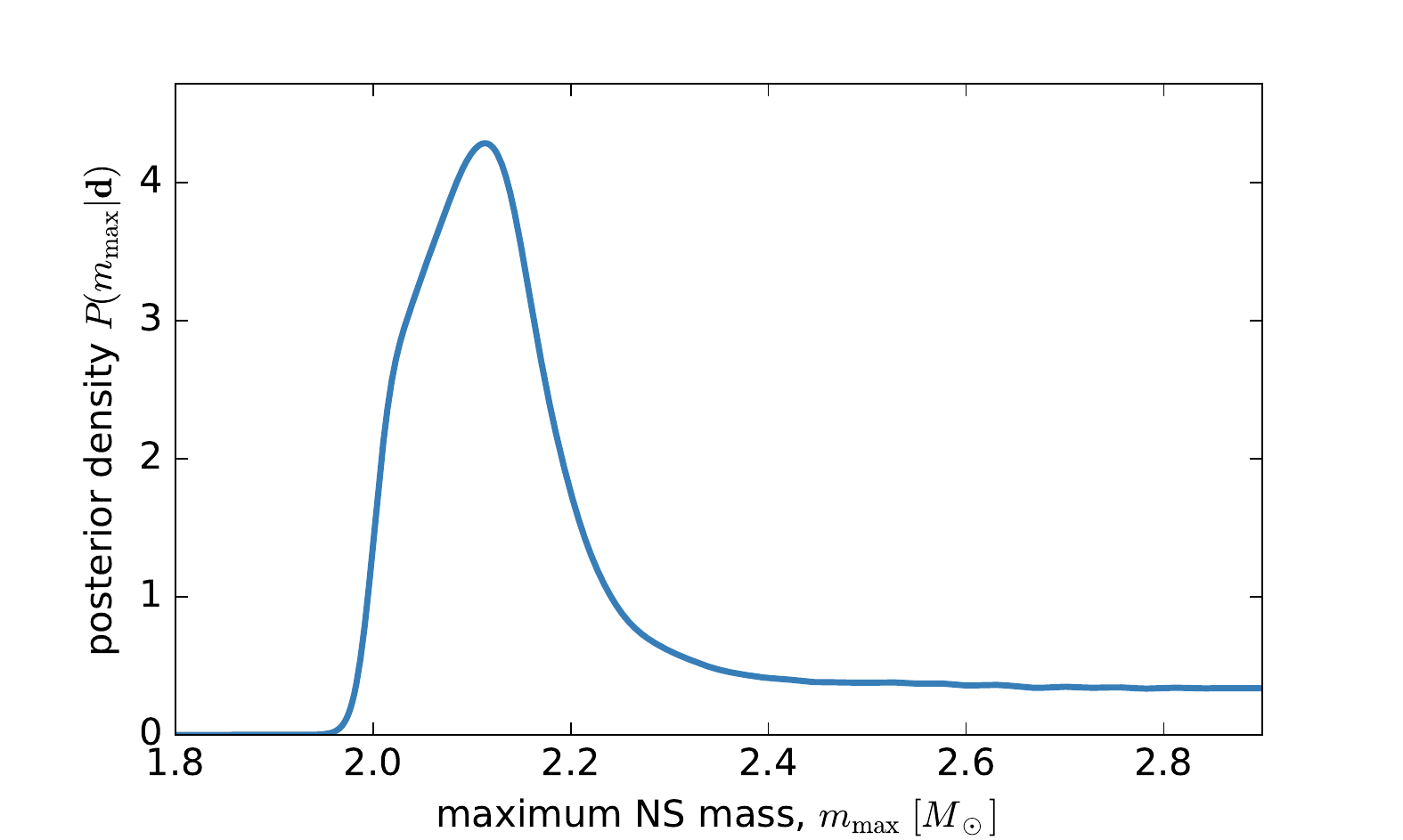}
\caption{Marginal posterior distribution for $\mmax$ derived from inferring the distribution of NS masses, assuming the mass-distribution can be modeled as the sum of two Gaussians with a hard cut at $\mmax$.}
\label{fig:posterior_mmax_fiducial}
\end{figure}

As discussed above, we find evidence for a sharp cut-off in the NS mass distribution under all models considered, with Bayes factors of $2\ln\mathcal{K}>3$ for all models $n\geq2$ (Table \ref{tab:evidences}). The marginal posterior distribution for the maximum NS mass under the fiducial $n=2$ model is shown in Fig. \ref{fig:posterior_mmax_fiducial}. The inferred posterior is peaked at $\mmax= 2.12\msol$, with credible regions $2.0\msol < \mmax < 2.2\msol$ (68\%) and $2.0 < \mmax < 2.6\msol$ (90\%)\footnote{Due to the fat tail of the posterior on $\mmax$, for the Bayesian credible regions we quote an iso-probability interval for the 68\% credible region, and a one-tail upper limit (with a hard lower limit) for the 90\% credible region.}. The lower limit on the maximum mass is hard, whilst the posterior has a fat (almost flat) tail out to $2.9\msol$; although the maximum mass has clearly been constrained, there persists a small but non-negligible possibility that $\mmax$ is still large.

Some constraints on $\mmax$ were reported in \citet{Antoniadis2016} who studied the distribution of millisecond pulsar masses using a similar (two component) Gaussian mixture model as used in this work. Our constraints are in good agreement with their results: cf. Fig. 10 of \citet{Antoniadis2016}. By considering all available NS mass data we are able to provide tighter constraints and, for the first time, substantial evidence that a sharp cut-off is preferred by the data; the mass cut-off is preferred with a Bayes factor of $2\ln\mathcal{K}>3$, whilst from the smaller subset of mass data considered in \citet{Antoniadis2016} the cut-off is only preferred with $2\ln\mathcal{K}=1.5$ (owing to the smaller sample). Nonetheless our results are entirely consistent with \citet{Antoniadis2016}.

Our constraints on the maximum NS mass are also in good agreement with, and independent of, recent studies of short GRBs, where \citet{Lawrence2015} and \citet{Fryer2015} argue that $\mmax \lesssim 2.2-2.5\msol$ is required assuming that the main source of short GRBs are NS-NS mergers. On the flip side, our constraints on the maximum NS mass show a strong preference for equations of state that produce short GRBs in NS-NS mergers. Combined with canonical values for binary NS merger rates \citep{Abadie2010, Dominik2015,Chruslinska2017}, our result strengthens the case for NS mergers as the primary source of short GRBs. Our results are also in good agreement with independent constraints on $\mmax$ from observations of the binary neutron-star merger GW170817 \citep{GW170817}, which give have been used to derive upper limits $\mmax < 2.17\msol$ (90\%) \citep{Margalit2017}, $\mmax < 2.33\msol$ (90\%) \citep{Rezzolla2018} and $\mmax < 2.16 - 2.28\msol$ \citep{Ruiz2018}.

In the following section we discuss the sensitivity of the inferred posterior on $\mmax$ to the choice of model and to key data cuts.
\begin{figure*}
\centering
\includegraphics[width = 17cm]{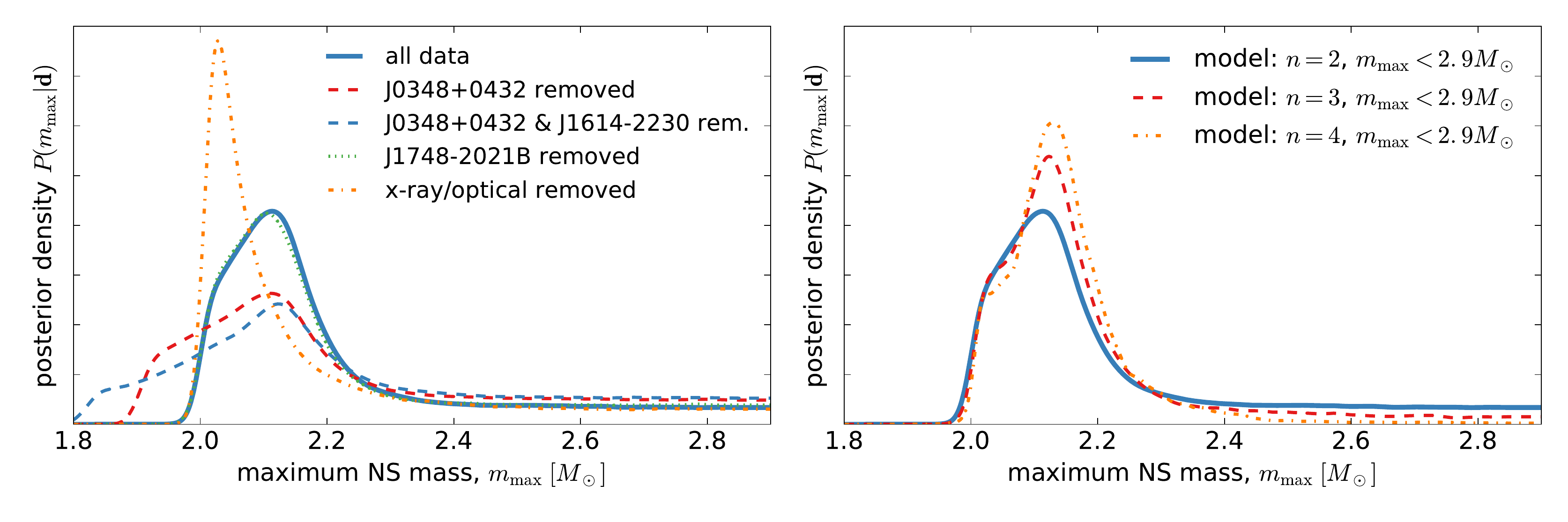}
\caption{Left: Sensitivity of the inferred marginal posterior distribution for $\mmax$ to removing key NSs from the dataset; keeping all NS mass data (blue-solid), removing J0348+0432 ($\mp=2.01(4)\msol$; red-dashed), removing J0348+0432 and also J1614-2230 ($\mp=1.93(2)$; blue-dashed), removing J1748-2021B ($\mp=2.74(21)\msol$; green-dashed) and removing the 21 x-ray/optical mass measurements (orange-dashed). Right: Sensitivity of the inferred marginal posterior distribution for $\mmax$ to the number of Gaussian components $n$ in the model: $n=2$ (fiducial; blue-solid), $n=3$ (red-dashed), $n=4$ (orange-dashed).}
\label{fig:posterior_mmax_models}
\end{figure*}
\subsubsection*{Sensitivity to data cuts}
\label{sec:cuts}
In Fig. \ref{fig:posterior_mmax_models} (left) we look at the impact of removing certain key NSs from the data. 

We should expect that the lower limit on $\mmax$ comes predominantly from the most massive precisely measured NSs available to date, namely J0348+0432 ($\mp=2.01(4)\msol$; \citet{Antoniadis2013}) and J1614-2230 ($\mp=1.93(2)$; \citet{Demorest2010, Fonseca2016}). From Fig. \ref{fig:posterior_mmax_models} (left; red and blue) it is evident that removing J0348+0432 or both J0348+0432 and J1614-2230 from the dataset impacts the lower limiit for $\mmax$, as expected, whilst otherwise preserving the qualitative shape of the posterior. Importantly, evidence for a cut-off in the mass distribution remains even with these important systems removed, albeit at a weaker level with $2\ln\mathcal{K} = 1.2$.

The other high-mass NS that one might suspect has a disproportionate impact on the inferred $\mmax$ is J1748-2021B\footnote{Other likely high-mass pulsars, such as the so-called black-widow pulsar B1957+20, are typically less extreme in their mass-likelihoods than J1748-2021B so should have an even smaller impact on the inferred $\mmax$.} \citep{Freire2008}. This system has a measured mass of $\mp=2.74(21)\msol$, as inferred from the total mass of that binary system (c.f. Eq. \eqref{total_mass_likelihood}). Although the uncertainty on its mass is substantial, since this system has almost all of its mass-likelihood at $\mp > 2\msol$ (with very low probability of being $< 2\msol$) and such a high peak value, this system could be strongly informative on the lower limit on $\mmax$ and responsible for the flat tail of the posterior density out to high values. From Fig. \ref{fig:posterior_mmax_models} (left; green) it is clear that removing this system has very little impact on the inference of $\mmax$.

The subset of the NS mass data that may be suspected to be contaminated by systematic errors, and hence biasing the inferred $\mmax$, are those mass measurements obtained through optical and x-ray observations \citep{Ozel2012, Falanga2015}. From Fig. \ref{fig:posterior_mmax_models} (left; orange) it iw clear that removing the x-ray/optical mass data has a much more substantial impact on the $\mmax$ posterior. This is hardly surprising, since the $21$ x-ray/optical masses removed represent a substantial fraction of the total sample of $74$, and they populate the higher-mass end of the distribution. Removing the x-ray/optical data shifts the posterior on $\mmax$ by around $0.1\msol$ and sharpens it (reducing the width at half-maximum by roughly a factor of two). The small shift and smearing of the posterior to higher $\mmax$ when including the x-ray/optical data makes good sense, since those systems occupy the higher mass end of the distribution and may hence encourage a slightly higher cut-off. Whilst it is possible that systematic biases in the x-ray/optical mass measurements bias the $\mmax$ inference, they would have to all be preferentially biased in the same direction for this effect to be significant, and even if this were the case we should expect the bias to be much smaller than the $0.1\msol$ shift induced by removing those data completely. Crucially, positive evidence for a sharp cut-off persists with the x-ray/optical masses removed -- the model with free $\mmax < 2.9\msol$ is preferred over $\mmax = 2.9\msol$ with a Bayes factor of $2\ln\mathcal{K} = 2.2$ (positive support for the cut-off). The fact that removing the x-ray/optical mass measurements does not significantly change our conclusions and any bias introduced by their systematics is expected to be small adds robustness to our results.

From these sensitivity tests, we conclude that the evidence for and inference of $\mmax$ is driven by the shape of the NS mass-distribution, informed by the whole population, rather than set exclusively by the most extreme objects observed (although the highest precicely measured masses dominate the lower limit on $\mmax$). Our key conclusion -- that there is evidence for a sharp cut-off in the NS mass distribution -- is robust to removing key subsets of the dataset.

\subsubsection*{Sensitivity to the mass distribution model}
\label{sec:models}
In Fig. \ref{fig:posterior_mmax_models} (right) we explore the sensitivity of the inference of $\mmax$ on the choice of model for the mass distribution. Clearly, the three Gaussian mixture models considered with $2$-, $3$- and $4$-components respectively yield a very similar posterior distribution for $\mmax$, with a small shift in the peak value and modestly tighter constraints on $\mmax$ as the number of components is increased. Note that the latter observation is also corroborated by the Bayesian evidences for the various models given in Table \ref{tab:evidences}, i.e., the evidence for a cut-off $\mmax < 2.9\msol$ versus $\mmax=2.9\msol$ increases with the number of components. The fact that the $\mmax$ posterior is so insensitive to the number of components in the model is also unsurprising in light of Fig. \ref{fig:mass_dist_fid}; all models give a similar bimodal distribution, where the key change is in the shape of the higher-mass mode but the qualitative characteristics of the distribution are very similar.

Whilst the sample size of measured NS masses is small and we restricted our analysis to the space of Gaussian mixture models, the lack of sensitivity to the model choice seen in Fig. \ref{fig:posterior_mmax_models} adds robustness to our results. 
\subsubsection*{Selection effects and accretion}
It is possible that selection effects may be influencing our inference of the mass distribution. The mass measurements used are exclusively from NSs in binaries, so our results should not be assumed to be applicable to isolated NSs. Furthermore, Shapiro delay detection (leading to the most precise mass measurements) is easier in systems with short orbital periods and higher inclinations, and spectroscopic observations are more relevant to relatively compact systems. Whilst it is possible that these selection effects may leave some imprint on the inferred mass distribution, it seems implausible that they are responsible for the inferred hard cut-off at $\mmax$.

The high-mass end of the distribution is a strong probe of both formation of higher mass NSs and accretion physics. In the absence of a hard cut-off due to the EoS, one would expect formation and accretion alone to generate a distribution with a smooth high-mass tail. The Gaussian mixture models used here are flexible enough to capture a smooth tail, even if it were skewed to give non-Gaussian (eg., steeper) dropoff; since the data preferred a sharp cut-off under our flexible model, we attribute the maximum mass cut-off to the EoS. Nevertheless, understanding the detailed formation and accretion physics and selection effects that underpin the inferred mass distribution is crucial to building confidence (or otherwise) in this conclusion.
\subsubsection*{What would happen if we observed a $2.1\msol$ neutron star?}
It is interesting to ask what would happen to our constraints on $\mmax$ if we measured (with good precision) a neutron star with around $2.1\msol$, i.e. around the peak of the $\mmax$ posterior (Fig. \ref{fig:posterior_mmax_fiducial}). We find that adding a sharply measured $2.1\msol$ NS into the mass dataset shifts the lower limit on $\mmax$ up to around $2.1\msol$ (as one would expect), but otherwise the shape of the $\mmax$ posterior, including the tail out to high masses, is largely unchanged. This makes good sense: in order to better constrain the location of the cut-off in the mass distribution from above, the high-mass end of the mass distribution needs to be better resolved. This can only be achieved with more measured masses around the high-mass end of the distribution.
\section{Constraints on the neutron star EoS from the inferred $\mmax$}
\label{sec:eos_constraints}
Every EoS has associated with it a maximum stable NS mass that can be determined by solving the Tolman-Oppenheimer-Volkoff (TOV) structure equations \citep{Tolman1939, Oppenheimer1939}. As such, observational bounds on the maximum NS mass provide corresponding constraints on the NS EoS. Recent observations of $2\msol$ NSs \citep{Antoniadis2013, Demorest2010, Fonseca2016} have already ruled out some EoS models, and the requirement to support $>2\msol$ NSs has become one of the key requirements for the nuclear EoS at ultra-high densities.

From our inference of $\mmax$, the posterior probability of a model EoS with a given $\mmax^\mathrm{EoS}$ is given by
\begin{align}
\label{eos_posterior}
P(\mathrm{EoS} |\data) = P(\mmax = \mmax^\mathrm{EoS} | \data),
\end{align}
where the right hand side is just the posterior density shown in Fig. \ref{fig:posterior_mmax_fiducial}.

Throughout this section we will compute maximum masses by solving the TOV equations for slowly rotating NSs. This is justified, since NSs in binaries are neither expected nor observed to have spin periods much less than $\lesssim 1.5\mathrm{ms}$ \citep{Chakrabarty2008,Papitto2014,Miller2015,Patruno2017}. At these rotation rates, corrections to the mass due to rotation are less than a few percent, and typically much less\footnote{Even at 
the extreme, mass-shedding limit, the mass of a nonrotating maximum mass neutron stars increases at most by $\sim 20\%$ due rotation~\citep{Breu:2016ufb}, this value being smaller for lower mass stars~\citep{Cook1994}. For reference, the fastest spinning pulsar in a binary, PSR J1748-2446ad \citep{Hessels2006} has a spin period of $\sim 1.4\mathrm{ms}$, a factor of $\sim 1.5$ larger than the mass-shedding frequency for a rotating NS~\citep{Lattimer2004} if we consider a fiducial nonrotating massive NS with mass $M = 2 M_{\odot}$ and radius $R = 14 \mathrm{km}$.} \citep{Cook1994,Stergioulas1995,Berti:2003nb,Berti:2004ny,Yagi:2014bxa}.

In the following we explore constraints on the most up-to-date tabulated EoS models. In \S \ref{sec:polytropic_eos} we constrain a parameterized piecewise-polytropic model for the EoS and obtain bounds on the maximum sound speed attained inside NSs.
\subsection{Constraints on numerical EoS models}
\label{sec:numerical_eos}
\begin{figure}
\centering
\includegraphics[width = 9.5cm]{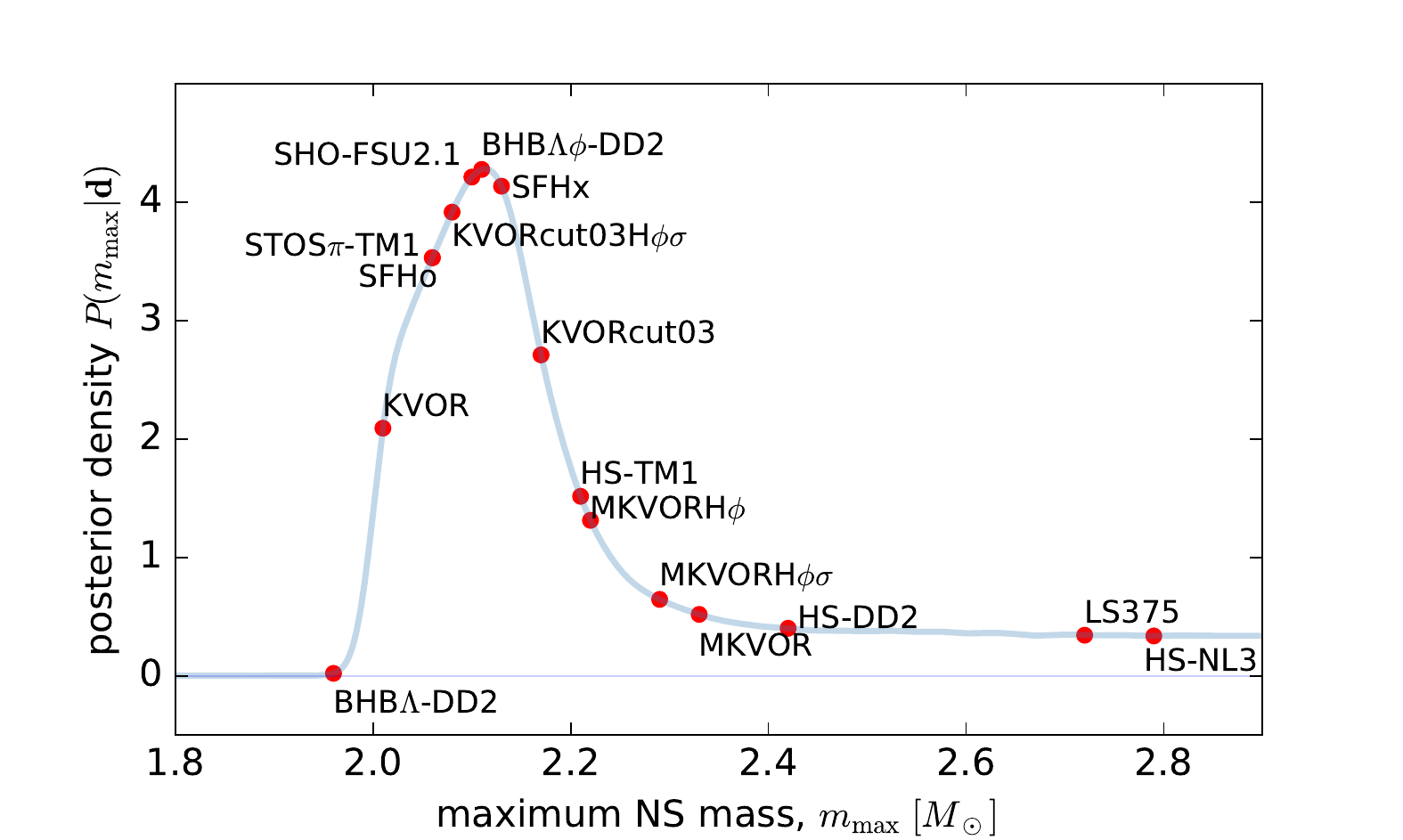}
\caption{Constraints on tabulated equations-of-state from the inferred posterior distribution of $\mmax$ derived from the NS mass distribution.}
\label{fig:tabulated_eos_constraints_current}
\end{figure}

In Table \ref{tab:eos} we list a selection of nuclear physics-based tabulated EoS models, along with their respective maximum NS masses. This set of EoS tables is not intended to be exhaustive: for a more complete catalog, see e.g. \citet{Oertel2017}. Fig. \ref{fig:tabulated_eos_constraints_current} shows the relative posterior probabilities of these EoSs given our posterior inference of $\mmax$ (also tabulated in Table \ref{tab:eos}), computed from Eq. \eqref{eos_posterior}. From Fig. \ref{fig:tabulated_eos_constraints_current} it is clear that our analysis of the NS mass distribution strongly favors some EoS models relative to others, at odds ratios of up to $12:1$ (see also Table \ref{tab:eos}). This is a vast improvement over previous maximum mass considerations, where any EoS that supported NSs with $\gtrsim 2\msol$ was considered equally acceptable based on maximum mass considerations alone.

Whilst caution is justified since our results are derived from a still relatively small sample of $74$ measured NS masses, the robustness of our results to model choice and data cuts demonstrated in \S \ref{sec:mmax_constraints} builds confidence in these constraints. Furthermore, our results fall nicely in line with independent constraints on the maximum NS mass derived from the assumption that short GRBs are produced primarily by coalescing NSs with rapidly collapsing remnants \citep{Lawrence2015, Fryer2015}. This begins to paint a coherent picture of upper limits on the maximum NS mass from astrophysical observations and considerations, and tightens our grip on $\mmax$.

%
%
\begin{table*}
\centering
\begin{tabularx}{\textwidth}{ccccc}
\toprule
EoS  & Exotic d.o.f & $\mmax$ & posterior probability & reference(s)\tabularnewline
\midrule
HS-DD2 				& 				&$2.42$ 	& 0.402 & \citet{Hempel2010, Fischer2014}	\\
SFHo 				&				&$2.06$ 	& 3.532 & \citet{Steiner2013}				\\
SFHx   				&				&$2.13$	& 4.136 & \citet{Steiner2013}\\
HS-NL3  				&		 		&$2.79$	& 0.337 & \citet{Hempel2010, Fischer2014}\\
SHO-FSU2.1  			& 				&$2.12$ 	& 4.262 &\citet{Shen2011} \\
LS375  				& 				&$2.72$	& 0.343 & \citet{Lattimer1991}\\
HS-TM1  				& 				&$2.21$	& 1.512 & \citet{Hempel2010, Hempel2012}\\
KVOR  				& 				&$2.01$	& 2.092 & \citet{Kolomeitsev2005, Klahn2006}\\
MKVOR 				& 				&$2.33$ 	& 0.519 & \citet{Maslov2016}\\
KVORcut03 			& 				&$2.17$	& 2.711 & \citet{Maslov2016}\\
KVORcut03H$\phi\sigma$ & H, $\phi$, H$\sigma$ scaling		&$2.08$ 	& 3.917 & \citet{Maslov2016}\\
MKVORH$\phi$ 	& H, $\phi$	&$2.22$ 	& 1.313 & \citet{Maslov2016}\\
MKVORH$\phi\sigma$ & H, $\phi$, H$\sigma$ scaling			&$2.29$ 	& 0.646 & \citet{Maslov2016}\\
BBH$\Lambda$-DD2 	& H 		&$1.96$ 	& 0.020 & \citet{Banik2014}				\\
BBH$\Lambda\phi$-DD2  	& H, $\phi$	&$2.11$ 	& 4.278 & \citet{Banik2014}				\\
STOS$\pi$-TM1  		& $\pi$				&$2.06$ 	& 3.532 &\citet{Nakazato2008}\\
\bottomrule
\end{tabularx}
\caption{Maximum NS mass predictions and their associated posterior probabilities $P(\mmax | \data)$ (c.f. Fig. \ref{fig:tabulated_eos_constraints_current}) for a selection of tabulated NS equations-of-state. Additional exotic species are denoted by H (Hyperons), $\pi$ (pions), $\phi$ ($\phi$-mesons) and $\sigma$ ($\sigma$-mesons).}
\label{tab:eos}
\end{table*}
\begin{figure*}
\centering
\includegraphics[width = 17.5cm]{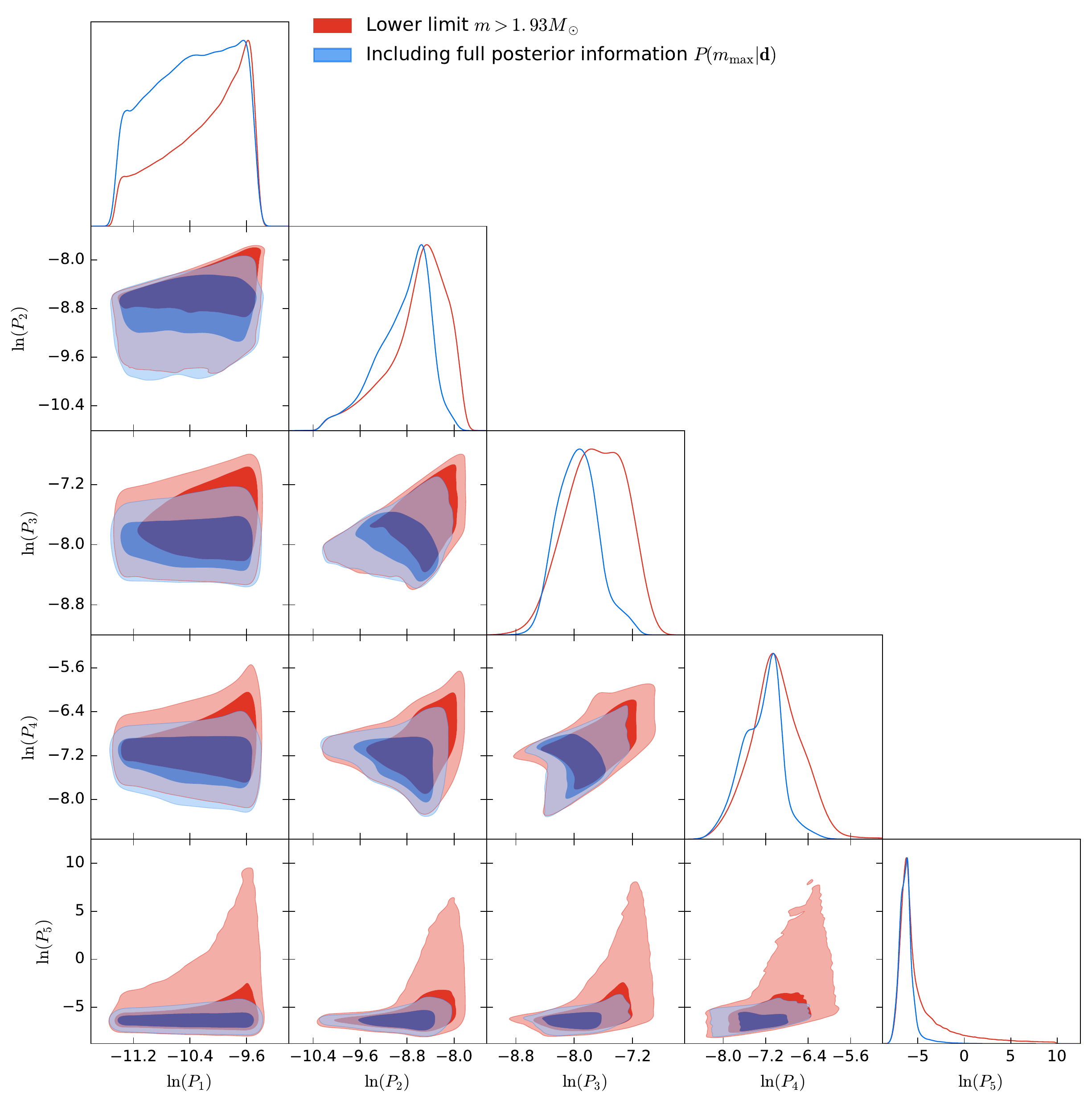}
\caption{Comparison of posterior constraints on the NS equation-of-state parameterized by a 5-piece polytrope from assuming a lower limit on the maximum NS mass $\mmax > 1.93 M_\odot$ (red) compared to including full posterior information on the maximum mass derived from the mass distribution of NSs $P(\mmax | d)$.}
\label{fig:effect_mmax_posterior}
\end{figure*}
\begin{figure}
\centering
\includegraphics[width = 9.5cm]{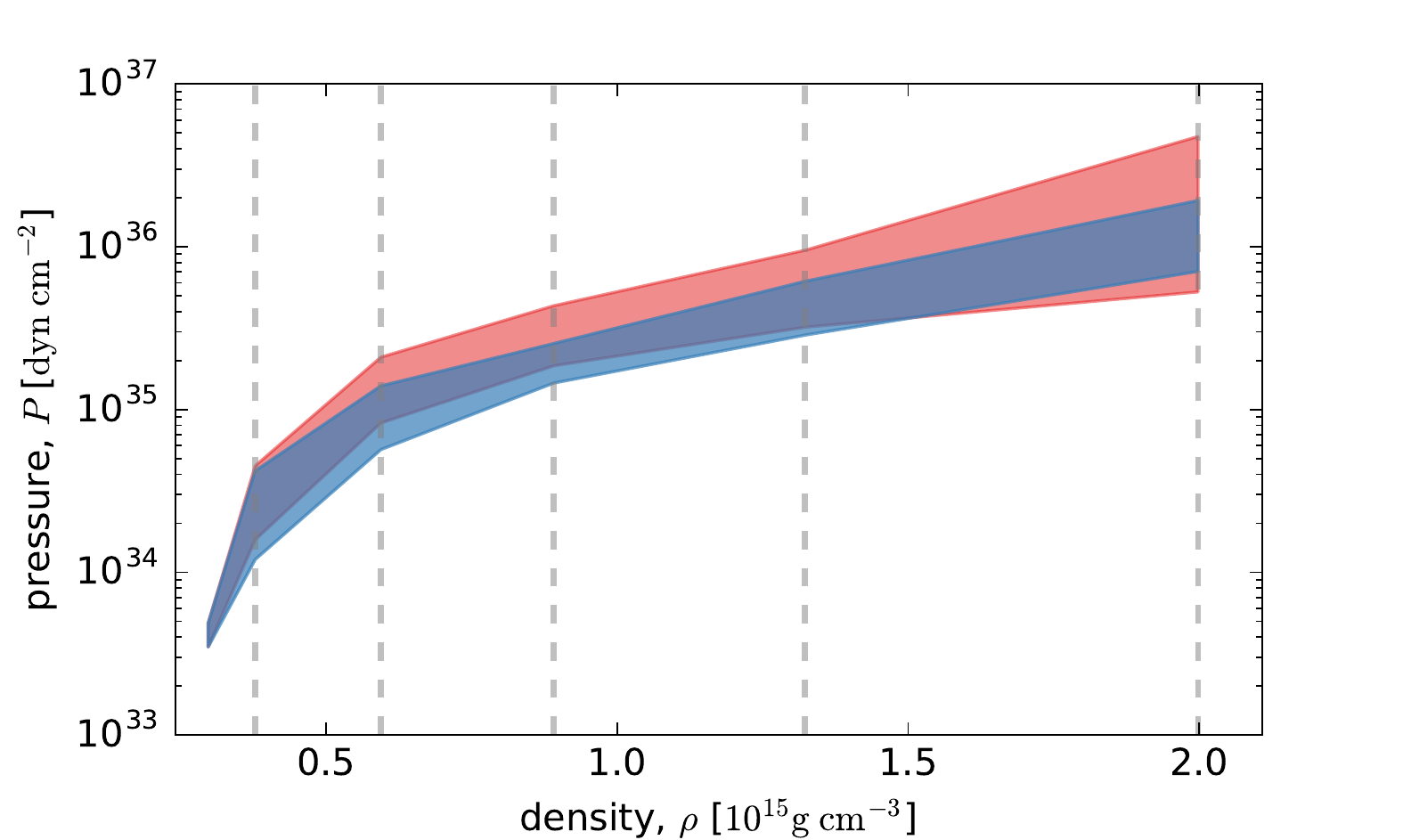}
\caption{Constraints on the piecewise polytropic EoS from the full posterior distribution for $\mmax$ derived from the NS mass distribution (blue), and just assuming a lower limit on the maximum NS mass $\mmax>1.93\msol$ (red). The bands indicate the 68\% credible regions for the pressure at the nodes $[1.4,\; 2.2,\; 3.3,\; 4.9,\;7.4]\rhoo$ indicated by the vertical grey lines. Note that the credible regions shown correspond to the 1d marginals; Fig. \ref{fig:effect_mmax_posterior} shows the correlation structure of the inferred polytropic EoS parameters. Assumptions about the EoS at densities $<1.1\rhoo$ (far left) are described in \S \ref{sec:polytropic_eos}.}
\label{fig:eos_band}
\end{figure}

\subsection{Constraints on piecewise polytropic EoS}
\label{sec:polytropic_eos}
\subsubsection{Piecewise polytropic model}
Representing the EoS as a piecewise polytrope has been demonstrated to be a useful parameterized model for the EoS\footnote{Other useful parameterizations also exist: for example \citet{Lindblom2010} performs a spectral decomposition of the EoS.} \citep{Read2009, Ozel2009, Hebeler2013, Steiner2016, Raithel2016}. Recently \citet{Raithel2016} showed that five polytropic nodes are required above the nuclear saturation density $\rhoo = 2.7\times 10^{14}\mathrm{g\,cm^{-3}}$ in order to reproduce the mass, radius, and moment of inertia for a range of realistic EoSs within the expected uncertainties of next-generation experiments (within $0.5$km, $0.1\msol$, and $10$\% respectively). In this work we follow \citet{Raithel2016} and construct a five-node piecewise-polytropic EoS above the saturation density, described by:
\begin{align}
P(\rho) = K_i\rho^{\Gamma_i},\;\;\;\;\rho_{i-1} < \rho < \rho_{i},
\end{align}
where the polytropic indices $\Gamma_i$ and normalization constants $K_i$ are determined by the pressures and densities at the knots,
\begin{align}
&\Gamma_i = \f{\ln(P_i/P_{i-1})}{\ln(\rho_i/\rho_{i-1})},\;\;\;\;K_i = \f{P_i}{\rho_i^{\Gamma_i}}.
\end{align}
Following \citet{Raithel2016}, we define five density knots at $[1.4,\; 2.2,\; 3.3,\; 4.9,\;7.4]\rhoo$, i.e. keeping the pressures at those densities $\mathbf{p} = (P_1, P_2, P_3, P_4,P_5)$ as our free model parameters of interest. 

Below $0.6\rhoo$ we assume a fixed SLy EoS for the crust taken from \cite{Douchin2001}. Between $0.6\rhoo$ and $1.1\rhoo$, the EoS is well constrained by chiral effective field theory calculations \citep{Tews2013, Hebeler2013, Kruger2013}. We assume the EoS in this regime is constrained to lie between the soft and stiff limits tabulated in \cite{Hebeler2013}; to implement this constraint, we introduce two additional polytropic knots at $\rho^*_1 = 0.6\rhoo$ and $\rho^*_2 = 1.1\rhoo$, where the pressures at those densities are tightly constrained to fall in the ranges $P^*_1\in [0.447, 0.696]\;\mathrm{MeV\;fm^{-3}}$ and $P^*_2\in [2.163, 3.542]\;\mathrm{MeV\;fm^{-3}}$ respectively (taken from Table 5 of \citet{Hebeler2013}).
\subsubsection{Physical constraints and priors}
\label{sec:polytrope_priors}
In addition to assumptions about the EoS below $1.1\rhoo$, we impose some additional physical constraints and priors at higher densities. We require that the EoS be microscopically stable, i.e. $P(\rho)$ must be strictly increasing:
\begin{align}
P_i > P_{i-1}.
\end{align}
We also require that the EoS does not violate causality, i.e. the local sound speed is smaller than the speed of light:
\begin{align}
\f{dP}{d\epsilon} = \f{c_s^2}{c^2} \leq 1.
\end{align}
Finally, some important constraints on the EoS at densities near $\rhoo$ come from nuclear scattering experiments at energies below $350 \mathrm{MeV}$ \citep{Ozel2016a, Raithel2017}; we impose lower limits on the pressures at the first two knots $P_1 \geq 3.60\;\mathrm{MeV\;fm^{-3}}$ and $P_2 \geq 11.70\;\mathrm{MeV\;fm^{-3}}$ to be consistent with nucleon-nucleon scattering data (following \citealt{Ozel2016a}; see also \citealt{Akmal1998, Pieper2001, Gandolfi2014, Raithel2017}).
\subsubsection{Bayesian inference of polytropic EoS in light of the measurement of $\mmax$}
As for the numerical EoS considered in \ref{sec:numerical_eos}, each piecewise-polytropic EoS $P(\rho;\mathbf{p})$ has associated with it a maximum stable NS mass $\mmax = \mmax(\mathbf{p})$, that can be found by integrating the TOV equations as before. We can hence use the likelihood for the maximum NS mass obtained in \S \ref{sec:mmax_constraints} to perform posterior inference for the polytropic EoS parameters $\mathbf{p}$:
\begin{align}
\label{polytrope_bayes}
P(\mathbf{p} | \data) &\propto P(\data | \mathbf{p})P(\mathbf{p}) \nonumber \\
&= P(\data | \mmax(\mathbf{p}))P(\mathbf{p}),
\end{align}
where we will assume broad log-uniform priors $P(\mathbf{p})$ additionally satisfying the constraints described in \ref{sec:polytrope_priors}, and the likelihood $P(\data | \mmax(\mathbf{p}))$ is given by Eq. \eqref{theta_posterior} marginalized over the other mass-distribution parameters (cf. Fig. \ref{fig:posterior_mmax_fiducial}).

To quantify the extra information gained from our results over previous NS mass considerations, we consider two cases: one using the full $P(\data | \mmax)$ obtained in \S \ref{sec:mmax_constraints}, and one just imposing $P(\data | \mmax) \propto \Theta(\mmax - 1.93\msol)$, i.e. assuming all of the information about the maximum mass comes from the most massive precisely measured system J0348+0432 ($\mp=2.01(4)\msol$; \citet{Antoniadis2013}) requiring $\mmax > 1.93\msol$ (95\%). We sample the posterior for the polytropic parameters using affine-invariant ensemble MCMC \textsc{emcee} \citep{Foreman2013}; the posteriors are summarized in Fig. \ref{fig:effect_mmax_posterior}.

From Fig. \ref{fig:effect_mmax_posterior} it's clear that including the full inference of the maximum mass from the NS mass distribution constrains the allowed parameter space of the piecewise polytrope EoS compared to simply taking a lower bound for $\mmax$, with the improvement being most significant for $P_3$, $P_4$ and $P_5$ at $\rho_3=3.3\rhoo$, $\rho_4 = 4.9\rhoo$ and $\rho_5=7.4\rhoo$ respectively. The 68\% credible regions of the marginal posteriors for $P_3$, $P_4$ and $P_5$ are improved by 35, 30 and 50\% respectively. Note that the tail out to high values of $P_5$ (particularly for the $\mmax>1.93\msol$ contours) are a result of the fact that when $P_1,\dots,P_4$ are sufficiently large, the maximum central density may be close to or smaller than $\rho_4$, in which cases $P_5$ is unconstrained. This is a weakness of the piecewise polytrope set-up for the EoS and may be alleviated by an alternative EoS parameterization.

Fig. \ref{fig:eos_band} shows the constraints on the piecewise polytropic EoS $P(\rho)$; the bands show the 68\% credible regions of the 1d marginal posteriors of the polytropic parameters using full posterior information on $\mmax$ (blue) versus just assuming a lower limit $\mmax > 1.93\msol$ (red). Whilst the correlations between the polytropic parameters are not captured by this figure, the $30$--$50$\% improvement on the inferred EoS at $3$--$7\times\rhoo$ is clearly shown.

\begin{figure}
\centering
\includegraphics[width = 9.5cm]{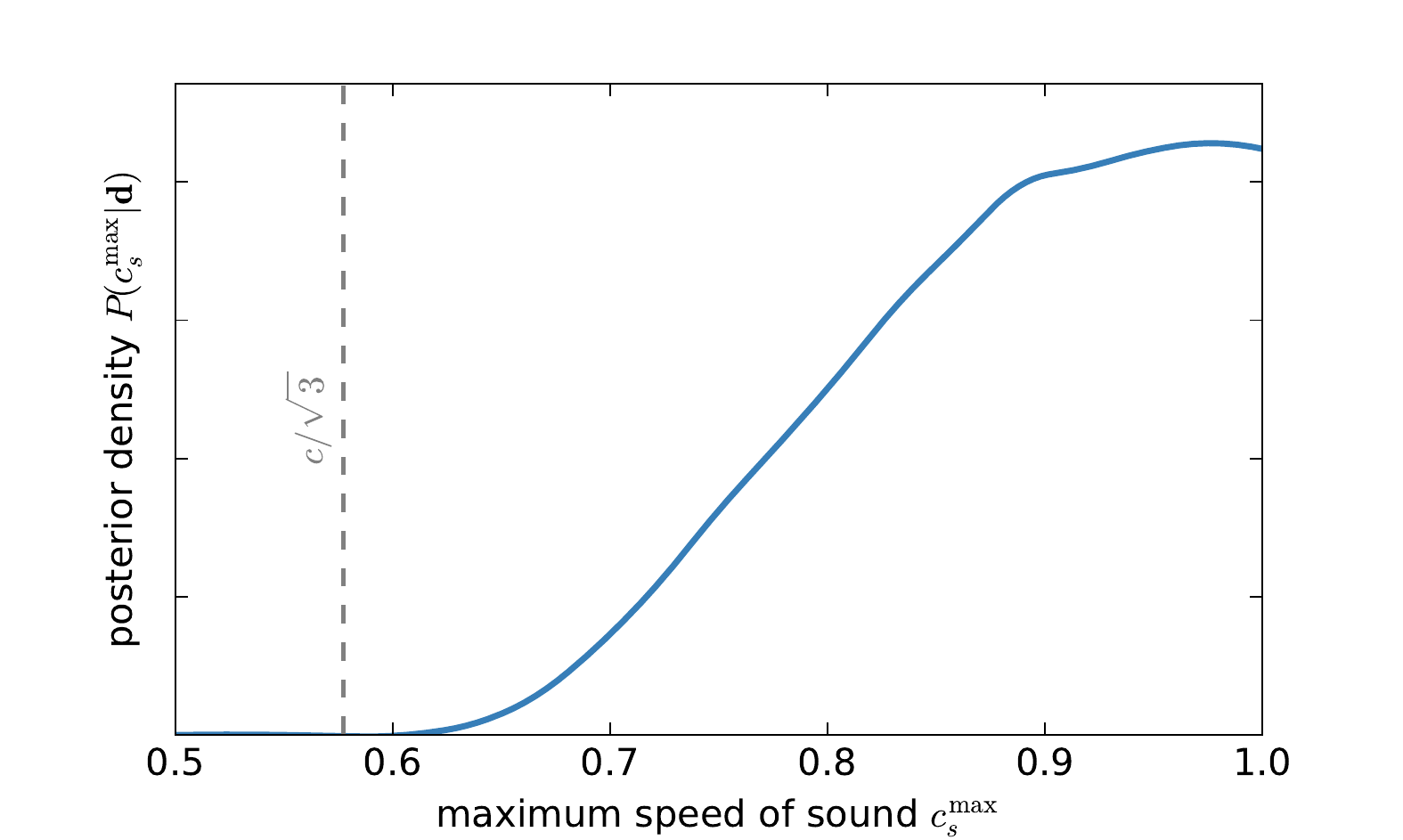}
\caption{Marginal posterior distribution for the maximum sound speed attained inside the NS, derived from $P(\mmax|\data)$ using the 5-piece polytropic model for the EoS. The lower limit is $c_s^\mathrm{max}>0.63c$ (99.8\%), ruling out $c_s < c/\sqrt{3}$ at high significance as shown by the vertical dashed line.}
\label{fig:sound_speed}
\end{figure}

There is an open debate on whether the causal limit $c_s < c$ is strict enough, with some authors making theoretical arguments why $c_s < c/\sqrt{3}$ may be a more appropriate bound \citep{Weinberg1972, Lattimer2014, Bedaque2015, Moustakidis2016}. In Fig. \ref{fig:sound_speed} we show the posterior constraints on the maximum speed of sound attained inside the NS under our piecewise polytropic model for the EoS. We find a lower bound on the maximum sound speed of $c_s^\mathrm{max} > 0.63c$ ($99.8\%$), ruling out $c_s < c/\sqrt{3}$ at very high significance. This tension was also discussed in \citet{Bedaque2015}. \citet{Kurkela2014} found that $c_s^\mathrm{max} > 0.74$ based on maximum mass considerations, although cautioned that their constraint may be overly restrictive.
\subsection{Comparison to other astrophysical constraints on the EoS}
In addition to maximum mass considerations, there are a number of other constraints on the NS EoS from astrophysical observations.

Measurements of NS masses and radii together can put strong constraints on the EoS \citep{Read2009,Ozel2009,Ozel2010,Steiner2010,Guillot2013,Steiner2013b,Steiner2013,Ozel2016a}. Measuring radii from x-ray observations is much more complicated than mass determination, and inferences are typically more model dependent; atmospheric composition, magnetic fields, source distance, interstellar extinction, residual accretion, brightness variations over the surface, and the effects of rotation in sources with unknown spin frequencies can all introduce systematic errors and must be carefully accounted for \citep{Miller2013,Potekhin2014,Fortin2015,Ozel2016a}. Whilst there is not yet firm consensus amongst NS radii measurements, advances have been made in accounting for systematic errors and strong (although model dependent) constraints on the EoS have been obtained (see eg., \citet{Ozel2016a} for an attempt to account for a multitude of systematics). Radius measurements are most sensitive to the EoS around $\sim2\times\rhoo$, so they are complementary to the maximum mass constraints, which probe higher densities $\gtrsim 3\times\rhoo$. 

\begin{figure}
\centering
\includegraphics[width = 9.5cm]{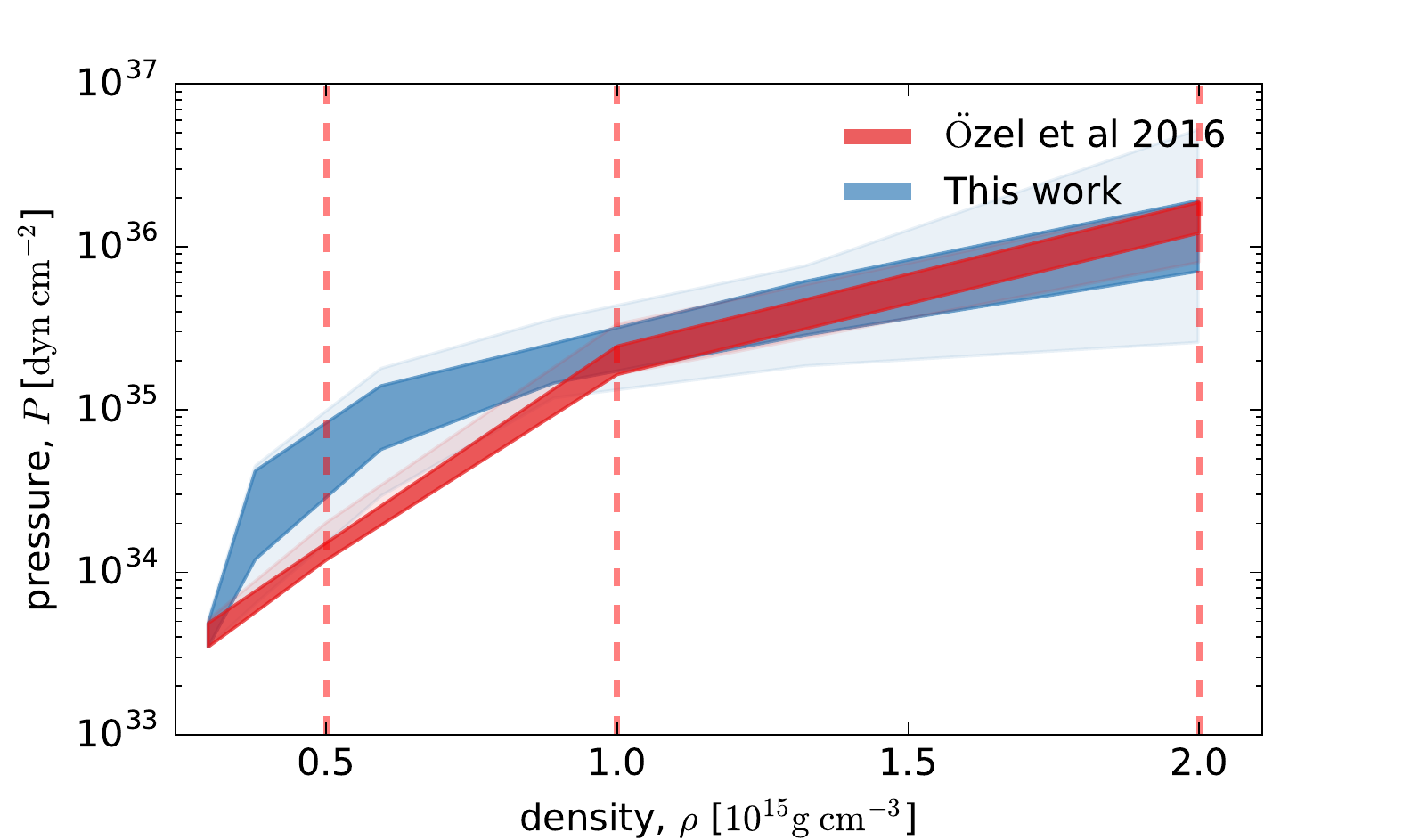}
\caption{Comparison of constraints on the piecewise polytropic EoS from the derived posterior distribution for $\mmax$ (this work; blue), and neutron star mass-radius measurements from \citealp{Ozel2016a} (red). The bands indicate the 68 and 95\% credible regions. The nodes of the five-node polytrope (this work) are at densities $[1.4,\; 2.2,\; 3.3,\; 4.9,\;7.4]\rhoo$, whilst \citet{Ozel2016a} used a three-node polytrope with nodes at $[1.85,\; 3.7,\; 7.4]\rhoo$, indicated by the red-dashed lines. Credible regions shown correspond to the 1d marginals; the correlation structure of the inferred EoS parameters is not captured by this plot (cf. Fig. \ref{fig:effect_mmax_posterior}). Priors and assumptions about the EoS at densities $<1.1\rhoo$ (far left) are similar in the two studies (see \S \ref{sec:polytropic_eos}).}
\label{fig:eos_band_comparison}
\end{figure}

In Fig. \ref{fig:eos_band_comparison} we compare the marginal constraints on the polytropic EoS from our work (blue) to the mass-radius measurement analysis of \citet{Ozel2016a} (red). Note that we use a more flexible five-node polytropic EoS, whereas \citet{Ozel2016a} use a three-node EoS with nodes at $[1.85,\; 3.7,\; 7.4]\rhoo$, but very similar prior assumptions otherwise (see \citealp{Ozel2016a} for details). As expected, the mass-radius measurements are more constraining than maximum-mass considerations alone, giving $\sim 4\times$ stronger constraints at $\sim 2\rhoo$ where the mass-radius data are most informative, and $\sim 2\times$ stronger constraints above $2\rhoo$. Although caution is deserved directly comparing the three- and five-node polytropic models, the figure gives some idea of the relative constraining power of these data under similar prior assumptions, and indicates statistical consistency between the two analyses.

\begin{figure}
\centering
\includegraphics[width = 9.5cm]{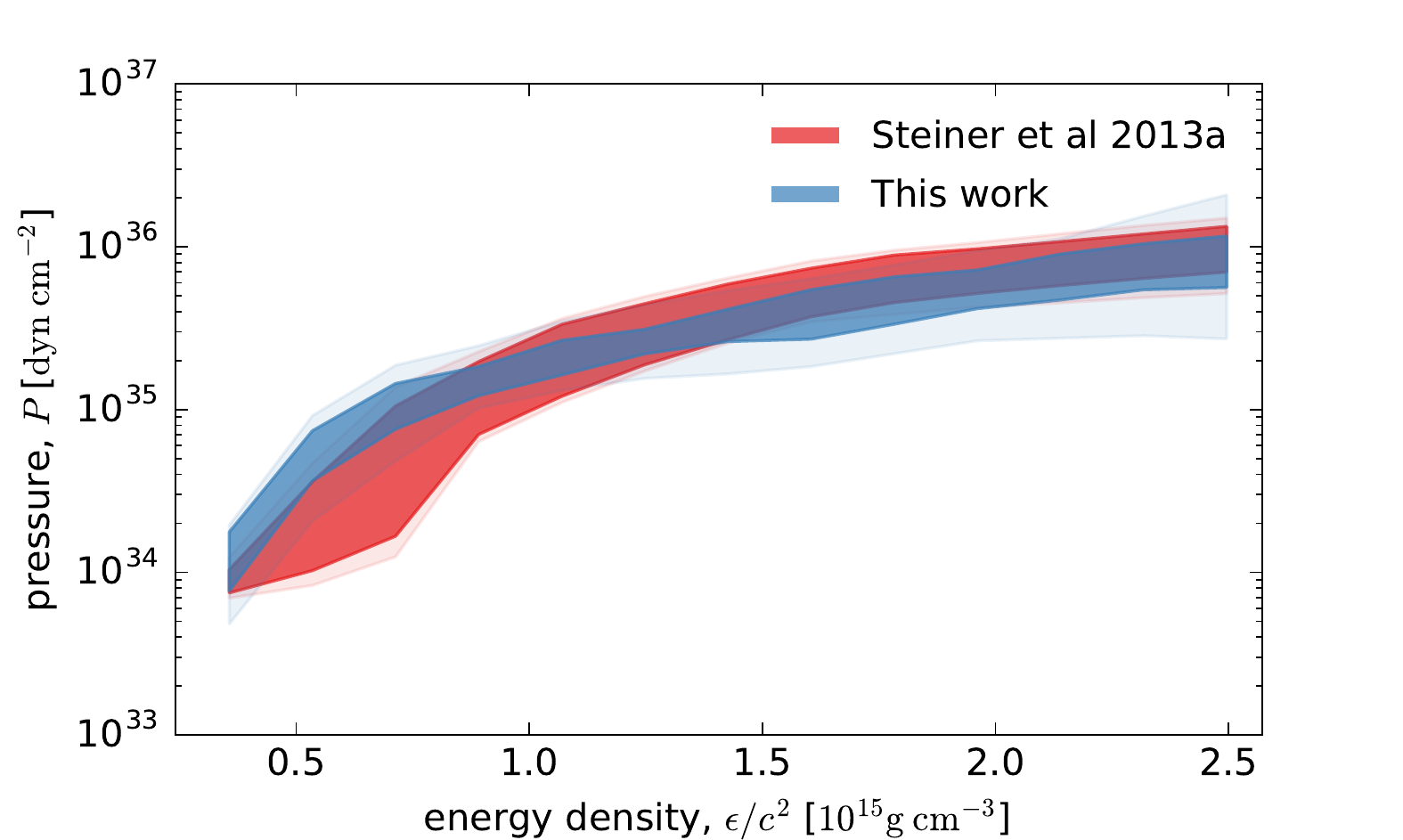}
\caption{Comparison of constraints on the EoS from the derived posterior distribution for $\mmax$ (this work; blue), and neutron star mass-radius measurements (\citealp{Steiner2013b}; red). The bands indicate the 68 and 95\% credible regions. Although \citet{Steiner2013b} consider a more flexible suite of EoS parameterizations compared to our five-node polytrope, the constraints are in good agreement. Note that our constraints on $P(\rho)$ (Figs. \ref{fig:effect_mmax_posterior}, \ref{fig:eos_band} and \ref{fig:eos_band_comparison}) were converted to $P(\epsilon)$ for direct comparison with \citet{Steiner2013b}.}
\label{fig:eos_band_comparison_steiner}
\end{figure}

In Fig. \ref{fig:eos_band_comparison_steiner} we compare our constraints on the EoS to the mass-radius measurement analysis of \citet{Steiner2013b}\footnote{The EoS constraints from \citet{Steiner2013b} are publicly available at \url{https://web.utk.edu/~asteine1/slb13.html}}. In order to compare with \citet{Steiner2013b}, we converted our constraints from $P(\rho)$ to $P(\epsilon)$ (pressure as a function of energy density), using the relation $d(\epsilon/\rho)=-P\,d(1/\rho)$. Since \citet{Steiner2013b} consider a more flexible suite of models for the EoS compared to our piecewise polytrope, this figure does not indicate the relative constraining power of the two datasets, but does demonstrate that our constraints are in good agreement with \citet{Steiner2013b}.

Gravitational wave observations of coalescing NS-NS or NS-black hole binaries with Advanced LIGO and VIRGO are expected to put constraints on the EoS primarily via determination of the NS tidal deformability \citep{DelPozzo2013,Agathos2015,Lackey2015}. The recent observation of the NS-NS merger GW170817 \citep{GW170817} marks the dawn of a new era in probing NS physics with gravitational waves, and it is already providing new insights on the EoS \citep{Annala2017,Banik2017,Radice2018,Most2018,Zhang2018}. We leave a direct comparison with these observations (and improved, combined constraints) to future work.

A measurement of the moment of inertia of NSs may be possible through observations of spin-orbit coupling effects in binaries, namely precession of the orbital plane and higher-order contributions to the periastron advance \citep{Lyne2004,Lattimer2005}. Such a measurement would provide further constraints on the EoS, particularly at $\sim2\times\rhoo$ \citep{Lattimer2005}, but sufficiently precise observations are expected to be some years away \citep{Lattimer2016}.

It has also been suggested that the lightest NSs may be used to constrain the EoS via their formation history \citep{Podsiadlowski2005}, although this approach typically leads to highly model dependent constraints. 
\section{Conclusions}
\label{sec:conclusions}
We have inferred the NS mass distribution from all currently available pulsar mass measurements using a flexible $n$-component Gaussian mixture model, allowing for a maximum mass cut-off. We find strong evidence for a bimodal distribution (Bayes factor $2\ln\mathcal{K}>10$), in agreement with previous literature. Increasing the number of Gaussian components in the mixture model does not elicit further distinct peaks, and the model with $n=2$ components is the most preferred.

We report, for the first time, positive evidence for a sharp cut-off in the NS mass distribution; for all Gaussian mixture models considered with $n\geq2$ components, the truncated models with $\mmax$ as a free parameter are preferred over those with fixed $\mmax=2.9\msol$ with Bayes factors $2\ln\mathcal{K}\gtrsim 3$ (Table \ref{tab:evidences}). We inferred the marginal posterior distribution for the maximum NS mass (Fig. \ref{fig:posterior_mmax_fiducial}) and obtained $2.0\msol < \mmax < 2.2\msol$ (68\%) and $2.0\msol < \mmax < 2.6\msol$ (90\%) credible regions for $\mmax$. These constraints on $\mmax$ are robust to the number of Gaussian components included in the mixture model for $n=2,\dots,4$, where increasing the number of components tightens the constraints and increases the Bayes factor modestly (Fig. \ref{fig:posterior_mmax_models}).

The evidence for and constraint on the $\mmax$ cut-off are robust against removing key subsets of the data, demonstrating that the observed maximum mass cut-off is driven by the shape of the NS mass distribution, informed by the whole population, rather than set by the most extreme (massive) objects (Fig. \ref{fig:posterior_mmax_models}). The lower bound on $\mmax$ is mostly set by the most massive precisely measured masses, namely J0348+0432 and J1614-2230 (both close to $2\msol$), but the reported positive evidence for the maximum mass cut-off persists with these systems removed. Similarly, J1748-2021B is likely to be even more massive (although more uncertain), but leaving it out of the dataset had a negligible impact on our conclusions. X-ray/optical mass measurements potentially suffer from residual systematics that could be biasing the $\mmax$ inference; we found that with these systems removed from the data, positive evidence for the sharp cut-off persists, and the inferred $\mmax$ was modestly impacted.

Our constraints on the maximum NS mass are in good agreement with, and independent of, recent studies of short GRBs, where \citet{Lawrence2015} and \citet{Fryer2015} argue that $\mmax \lesssim 2.2-2.5\msol$ is required assuming that the main source of short GRBs are NS-NS mergers. On the flip side, our constraints on the maximum NS mass show a strong preference for equations of state for which short GRBs are produced in NS-NS mergers, strengthening the case for NS mergers as the primary source of short GRBs. Our constraints on $\mmax$ are also in excellent agreement with the observations of the binary neutron star merger GW170817 \citep{GW170817}, which have been used to constrain $\mmax < 2.17\msol$ (90\%) \citep{Margalit2017}, $\mmax < 2.33\msol$ (90\%) \citep{Rezzolla2018}, $\mmax < 2.16 - 2.28\msol$ \citep{Ruiz2018}.

Using our inference of the maximum NS mass we are able to put tight constraints on the NS equation of state. We find that for a set of realistic EoSs that support $>2\msol$ NSs, our inference of $\mmax$ is able to distinguish between models at odds ratios of up to 12:1 based on maximum mass considerations alone (Fig. \ref{fig:tabulated_eos_constraints_current}). Considering a parameterized five-node polytropic equation of state, we are able to obtain constraints on the pressure at densities of $3$--$7\times\rhoo$ that are improved by $30$--$50\%$ compared to simply assuming $\mmax > 1.93\msol$ (Figs. \ref{fig:effect_mmax_posterior}-\ref{fig:eos_band}). Under this piecewise polytropic EoS model, we find a lower bound on the maximum sound speed attained inside the NS of $c_s^\mathrm{max}>0.63c$ ($99.8\%$), ruling out $c_s < c/\sqrt{3}$ at high significance (Fig. \ref{fig:sound_speed}). Our constraints on the EoS from maximum mass considerations are in good agreement with neutron star mass-radius measurements \citep{Steiner2013b,Ozel2016a}.

\section*{acknowledgments}
We thank Feryal \"{O}zel, Kenta Hotokezaka, Stephen Feeney, Paulo Freire and John Antoniadis for useful discussions, and Feryal \"{O}zel for kindly providing the data used for comparison in Fig.~\ref{fig:eos_band_comparison}. E.~B. was supported by NSF Grants No.~PHY-1607130 and AST-1716715, and by FCT contract IF/00797/2014/CP1214/CT0012 under the IF2014 Programme. H.~O.~S was supported by NSF Grant No. PHY-1607130 and NASA grant NNX16AB98G and 80NSSC17M0041. H.~O.~S also thanks Thomas Sotiriou and the University of Nottingham for hospitality.
\bibliographystyle{mnras_tex}
\bibliography{max_mass_eos}

\end{document}